\documentclass[journal]{IEEEtran}
\ifCLASSINFOpdf
\else
\fi

 \usepackage{amssymb,amsmath,epsfig,cite,url,multicol,multirow}
\usepackage{hyperref}
\usepackage{graphicx}
\usepackage{float}
\usepackage[hyperpageref]{backref}

\begin{document}
%
% paper title
% Titles are generally capitalized except for words such as a, an, and, as,
% at, but, by, for, in, nor, of, on, or, the, to and up, which are usually
% not capitalized unless they are the first or last word of the title.
% Linebreaks \\ can be used within to get better formatting as desired.
% Do not put math or special symbols in the title.
\title{Sonority Measurement using System, Source and Suprasegmental Information}
%
%
% author names and IEEE memberships
% note positions of commas and nonbreaking spaces ( ~ ) LaTeX will not break
% a structure at a ~ so this keeps an author's name from being broken across
% two lines.
% use \thanks{} to gain access to the first footnote area
% a separate \thanks must be used for each paragraph as LaTeX2e's \thanks
% was not built to handle multiple paragraphs
%

\author{Bidisha Sharma%,~\IEEEmembership{Student Member,~IEEE,}
        ~and~S. R. Mahadeva Prasanna,~\IEEEmembership{Senior Member,~IEEE}
        
\thanks{Bidisha Sharma and S. R. Mahadeva Prasanna are with the Dept. of Electronics and Electrical Engineering,
  Indian Institute of Technology Guwahati, Guwahati-781039, India. This work is part of ongoing project on development of text to speech synthesis systems in Indian languages.
Email: \{s.bidisha, prasanna\}@iitg.ernet.in}}% <-this % stops a space
\maketitle

\begin{abstract}
Sonorant sounds are characterized by regions with prominent formant structure, high energy and high degree of periodicity. In this work, the vocal-tract system, excitation source and suprasegmental features derived from the speech signal are analyzed to measure the sonority information present in each of them. Vocal-tract system information is extracted from the Hilbert envelope of numerator of group delay function. It is derived from zero time windowed speech signal that provides better resolution of the formants. A five-dimensional feature set is computed from the estimated formants to measure the prominence of the spectral peaks. A feature representing strength of excitation is derived from the Hilbert envelope of linear prediction residual,
which represents the source information. Correlation of speech over ten consecutive pitch periods is used as the suprasegmental feature 
representing periodicity information. The combination of evidences from the three different aspects of speech provides
better discrimination among different sonorant classes, compared to the baseline MFCC features. The usefulness of the proposed sonority feature is demonstrated in the tasks of phoneme recognition and sonorant classification.
\end{abstract}

% Note that keywords are not normally used for peerreview papers.

\vspace{-0.3cm}
\section{Introduction} \label{introduction}

%%Importance

   Sonority refers to relative loudness of speech sounds~\cite{parker2002quantifying}. Most of the sonorant sounds are produced using
   relatively less constricted vocal-tract shape and  glottal vibration. This results in regions of regular structure having
   high energy and high degree of periodicity. The sonorant regions are therefore prominent ones in the speech signal and important for many speech processing tasks~\cite{schutte2005robust}. Vowels are the most sonorous sounds, which mostly form the
   nucleus of a syllable. Different sonority hierarchies are defined in the literature as mentioned in~\cite{parker2002quantifying}. However, the most commonly referred sonority hierarchy for the six major classes of sonorants in the decreasing order of sonority is {\it low-vowels, mid-vowels, high-vowels, glides, liquids and nasals}. In~\cite{parker2008sound}, the sonority hierarchy for obstruents is defined in the decreasing order of sonority as {\it voiced fricatives, voiced affricates, voiced stops, voiceless fricatives, voiceless
   affricates, and voiceless stops}.
   
Sonority is used to explain both the perception of syllables and their phonetic structure~\cite{international1999handbook}. The {\it sonority sequencing principle} states that in every syllable, syllable nucleus has the highest sonority value~\cite{blevins2001syllable}. According to {\it syllable  contact law}, the junction between two syllables is well recognizable when the coda of the present syllable has higher sonority value than the onset of the next syllable~\cite{gouskova2004relational}. According to~\cite{de2004markedness}, the syllables with nuclei having more sonority value tend to have more stress compared to the syllables with nuclei having less sonority value. For example, syllables with [e] or [o] may be perceived as having more stress than those with [i] and [u]. The possible sequence of consonants present in the syllable onset and coda also depends on the sonority value associated with them. For example, consonant clusters present in syllable onset of the form [pl], [dr], [km] are very common,
 but the reverse order is rare. In this case, [l], [r], [m] are more sonorous than [p], [d], [k]. Similarly,  [mp] and [nd] are very common as syllable codas than [pm], [dn], where [m], [n] are more sonorous than [p], [d]. Therefore, sonority of a sound unit has an impact on the basic production pattern of speech sounds. In several studies of phonology such as consonant cluster, sonorant-obstruent cluster, syllable onset and coda position, degree of sonority is used~\cite{moreton2005syllabification,topbas2008reviewing}. {\it Degree of sonority} can be defined as sequential variation in various attributes that correlate to sonority, with respect to distinctive category of sound units. The variation in degree of sonority associated with different sound units is due to the change in the behavior of different articulators during production. This is also manifested in the produced speech signal.
\vspace{-0.3cm}
\subsection{Production aspects of different sonorant sounds}
% Sonority associated with a sound unit or speech segment depends on the configuration of different articulators in speech production mechanism. 
The most sonorant sounds, vowels, are produced with less constricted vocal-tract configuration through manipulation of the vocal-tract between glottis and lips. Position and configuration of different articulators has effect on the spectrum of generated speech signal. Narrowing the cross sectional area in the front part of vocal-tract and widening towards the back results in the decrease of  first formant frequency ($F_1$). As a consequence of variation in position and length of constriction, second formant frequency ($F_2$) changes for different category of sonorants. The bandwidth of formant is associated with loss in the vocal-tract. Thus with the increase in sonority, the vocal-tract constriction decreases that results in increase in $F_1$, $F_2$ and decrease in formant bandwidth. 

Compared to the obstruents, sonorants have sufficient opening of the vocal-tract to produce voicing and well defined prominent formant structure~\cite{beckman1992prosodic}. Looking into these aspects of sonorant sounds, it is expected that, accurately estimating the vocal-tract spectrum (VTS) and analyzing the formant structure may be helpful to characterize the change in vocal-tract shape with the change in degree of sonority. Due to the glottal open and closed phase, the formant structure does not show a constant behavior during one pitch period~\cite{ananthapadmanabha1982calculation,childers1994measuring}. The characteristics of the vocal-tract system in the open phase varies due to the coupling with vocal-fold and trachea. Whereas, during the closed phase, the speech signal is mainly due to free resonances since there is no coupling with trachea and vocal-folds~\cite{yegnanarayana1998extraction}. Therefore, extraction of VTS from speech signal corresponding to the closed phase of each pitch period may 
give accurate formant estimation along with it's associated measures. But, in voiced region, the glottal closing is abrupt and the duration of the closed phase is smaller than that of the open phase. For extracting the VTS, processes like linear prediction (LP) analysis and short time Fourier transform (STFT) involve block processing and are dependent on the size and position of window. Also, these methods mask the changing shape of the vocal-tract and give an average spectrum~\cite{yegnanarayana1998extraction}. 
         
In this work, Hilbert envelope of numerator of group delay function (HNGD) spectrum derived from speech signal around the glottal closure instant (GCI) is used to estimate the VTS~\cite{bayya2013spectro}. The GCI locations are estimated using the zero frequency filtered (ZFF) signal~\cite{murty2008epoch}, as it is found to be more robust compared to other state-of-the-art techniques~\cite{drugman2012detection}. A highly tapering window is used to emphasize the speech samples around each GCI that correspond to the glottal closed phase. The sonority information present in the VTS is extracted using knowledge from the first three formants of the HNGD spectrum.

With change in the vocal-tract constriction, there is also an effect on the amplitude and spectrum of the source. Due to the change in constriction, there is fluctuation in supra-glottal pressure which has an impact on the pressure inside the glottis during the open phase of glottal vibration. This changes mechanical motion of the vocal-folds. The net effect is reduction in the amplitude of glottal source which is reflected in the  Hilbert envelope (HE) of LP residual as strong peaks. These peaks have correlation with an acoustic feature called strength of excitation (SoE) as discussed in~\cite{seshadri2009perceived}. With the increase in degree of sonority, SoE also increases. Hence, it can be hypothesized that, deriving an adequate  representation of SoE may add some advantage in deriving sonority information from the speech signal. 

Along with the change in behavior of the vocal-tract system and the excitation source with degree of sonority, temporal variation in  the speech signal also takes place. This can be  observed over several pitch periods. One such measure is periodicity, which is tendency of the signal to repeat similar structure over several pitch periods. This occurs, since human speech production system changes in a continuous manner. During the production of sonorant sounds, the vocal-tract shape changes slowly and hence maintains periodicity over longer duration compared to other sounds~\cite{puppel1992sonority}. This suprasegmental behavior of sonorants is not taken into account while analyzing vocal-tract system and excitation source perspectives. Hence, examining the regularity in the signal structure or correlation over several small segments of the speech signal may be helpful to obtain feature representing this aspect of sonority. 
\vspace{-0.3cm}
\subsection{Usefulness of sonority feature}
 Deriving sonority feature from speech signal may be helpful in many speech processing tasks. These include, but are not limited to detection of syllable nucleus, vowel onset point detection, phoneme classification, study of syllable structure and syllabification in different languages. Sounds with higher degree of sonority form syllable nucleus. It gives information about number of syllables present in the speech signal. Number of syllables divided by duration of the signal defines syllable rate/speaking rate. There are several approaches in the literature towards this direction. In~\cite{pfitzinger1996syllable}, syllable nucleus is detected by loudness estimation. Energy peaks in the frequency range from 250 - 2500 Hz have good correlation with syllable nuclei. Many other methods use vowel recognizer to find syllable nucleus as given in~\cite{yuan2010robust}. 
 
Correlation between prominent subbands is used to capture well defined formant structure in the syllable nuclei in~\cite{wang2007robust}. Before applying cross-correlation between subband energy vectors, frames are weighted by Gaussian window and then temporal correlation is estimated in order to retain inter-syllable discontinuity in case of fast speech. Then, thresholding and pitch validation of subband correlation envelope is performed to enhance the detection of syllable nucleus. In the same work, experiments are also performed to find syllable nuclei which include sonorant sounds other than vowels. The mean error calculated is more in this case. This proves that the feature cannot detect all sonorant sounds. In~\cite{arrabothu2013syllable}, perceptually significant evidences such as excitation source peaks in LP residual and formant peaks which contribute to the loudness are used to find the most sonorous region within syllable. All these efforts are aimed to detect basically the most sonorous sounds, 
the vowels. There are many confusions reported within the sonorants (vowels, glides, liquids, nasals) while detecting the vowels. 
	
Segmentation of speech into sonorant regions with high accuracy is essential for applications like automatic speech recognition (ASR) to detect the regions with high signal to noise ratio (SNR) in the speech signal~\cite{dumpala2015robust}. In literature, sonorant segmentation is performed by using mel frequency cepstral coefficients (MFCCs), knowledge based acoustic features or a combination of both~\cite{schutte2005robust,jansen2008modeling}. Recently in~\cite{dumpala2015robust,sarma2015exploration}, features based on both spectral and source information are proposed and a hierarchical algorithm is developed to detect sonorant and non-sonorant regions in continuous speech. However, the feature may not have potential to further divide the sonorant regions based on the degree of sonority associated with the sound. In order to improve the performance of sonority detection, it is important to first quantify the degree of sonority associated with different sound units in a given speech segment, without having knowledge of phone 
sequence. In this work, an evidence is obtained which represents instantaneous sonority i.e. continuous change in sonority with time in the speech signal. In traditional methods, sonority is derived from the phone identity knowledge.
     
Looking into these studies present in the literature, it can be considered important to derive some feature which represents degree of sonority from speech signal. In this work, three different aspects of speech signal, namely vocal-tract system, excitation source and suprasegmental are analyzed to extract prospective features to discriminate among different classes of sonorants. The three attributes are analyzed individually and effectively combined to derive a multi-dimensional feature which can represent sonority. The obtained sonority feature is used in phoneme recognition and results show improvement. In the analysis of all features, focus is on classifying within the sonorants according to the sonority hierarchy.
 
Rest of the paper is organized as follows: Features of vocal-tract system for sonority detection are proposed in Section~\ref{system_feature}. Features of excitation source and suprasegmental feature are presented in Section~\ref{source_feature} and Section~\ref{suprasegmental_feature}, respectively. Section~\ref{combination} describes the combination of proposed evidences to represent sonority measure. Section~\ref{evaluation} shows the experiments performed to demonstrate the usefulness of sonority evidence in different speech processing task such as phoneme classifier. In Section~\ref{conclusion}, summary, conclusions and future direction are mentioned. 

\vspace{-0.3cm}      
 \section{Features of vocal-tract system for Sonority Detection} \label{system_feature}
The categorical formant structure in the VTS of sonorant sounds can be interpreted by measures associated with amplitude of spectral peaks and valleys, formant bandwidths and slope. Bandwidth of the spectral peak decreases, while the spectral peak value increases with increase in degree of sonority. The peak-to-valley ratio ($PVR$) of spectral peak is a direct representation of spectral prominence, that is inversely proportional to the corresponding bandwidth. Spectral prominence refers to spectral peaks with more sharpness and higher energy, which increases with degree of sonority. This depends on $PVR$, slope, bandwidth and amplitude associated with spectral peaks. Narrow constriction results in relatively low values of formant frequencies and spectral peaks. High-vowels are produced by raising the tongue body thus forming narrow constriction in the front part of vocal-tract. This results in decrease in $F_1$ and increase in bandwidth, primarily due to acoustic losses in the vocal-tract walls and glottis. 
As explained in~\cite{stevens2000acoustic}, due to less spacing between $F_1$ and $F_0$, the response of low frequency auditory nerve fibers are dominated in low frequency region by $F_1$, resulting in production of relatively stable response in auditory system. In contrast to high-vowels, low-vowels are produced by narrowing the posterior part and widening towards lips, resulting in increase in $F_1$ and higher difference between $F_1$ and $F_0$. Due to this difference, the auditory nerve fibers near $F_0$ are not dominated by $F_1$. As a consequence, there is a fall in the spectrum below $F_1$~\cite{stevens2000acoustic}. Due to the intermediate position of tongue body during production of mid-vowels, $F_1$ also lies in between that of high-vowel and low-vowel. In this case, the auditory nerve fibers are in synchrony with either $F_1$ or $F_0$. Fluctuation of second and third formant frequencies, $F_2$ and $F_3$ depends on the constriction length and position in the vocal-tract. 

During the production of nasals, the vocal-tract is completely closed, while the velopharyngeal part is open and there is no pressure increase behind the constriction. In this case, during the time of closure of vocal-tract, if the vocal-folds are in a position of voicing, the same will continue after the closure~\cite{stevens2000acoustic,sharma2016speech}. Nasals have the first formant at a very low frequency and with less energy. The higher formants are also of weak amplitudes. Glides are produced by forming narrow constriction to an extent, so that there is no significant pressure drop across the constriction. This results in vibration of vocal-folds and lower $F_1$ with wider bandwidth. As an influence of the narrow constriction, the glottal source also gets modified. The liquids are also produced with narrow vocal-tract constriction, but the length of the constriction is shorter than that of the glides. As a consequence, $F_1$ of liquids is higher than that of glides. During production of liquids, the tongue is shaped 
in such a way that there is a split in the vocal-tract, which cannot be compared with an uniform tube~\cite{stevens2000acoustic}. 

With the increase in vocal-tract constriction, $F_1$ decreases and bandwidth of first formant increases gradually along the sequence of following sounds: low-vowels, mid-vowels, high-vowels, liquids, glides and nasals. With decrease in $F_1$, there is significant reduction in the overall spectrum amplitude. Amplitude of $F_2$ is dependent on $F_1$ and on the point of constriction along the vocal-tract. Since sonority associated with a sound unit depends on the vocal-tract constriction, the process for extraction of VTS should be appropriate.
   
 \subsection{HNGD Spectrum}  \label{Zero Time Windowing}

        \begin{figure*}[t]
	\centering
	\centerline{\epsfig{figure=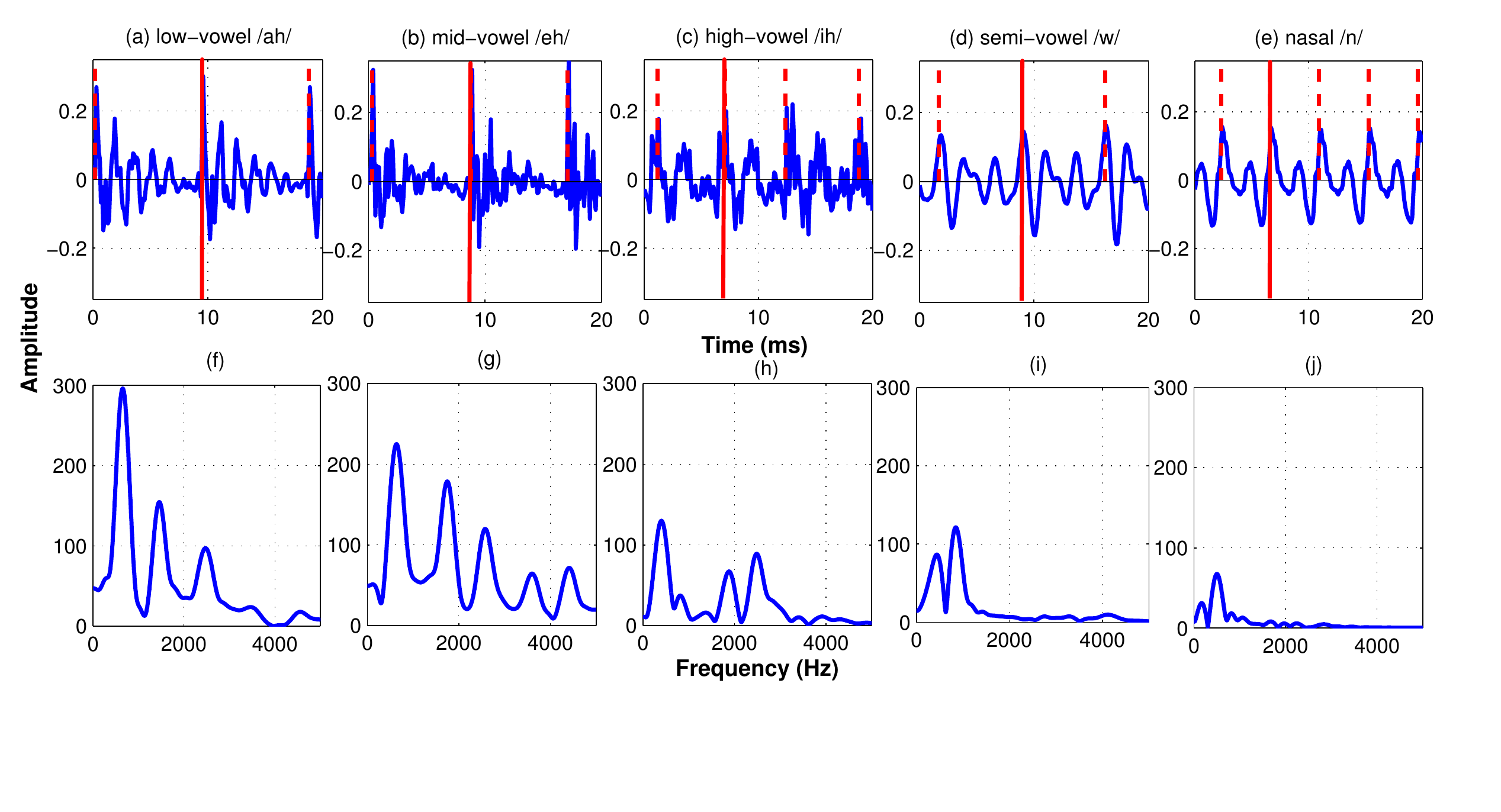,height=2.7in, width=6.2in}}
	\vspace{-1.2cm}
	\caption{{\it  HNGD spectra for different classes of sounds showing apparent discrepancy in the spectrum shape. First row depicts $20$ ms segment of (a) low-vowel /ah/, (b) mid-vowel /eh/, (c) high-vowel /ih/, (d) semi-vowel /w/, (e) nasal /n/ from TIMIT test database with dashed vertical lines representing epoch locations. Second row (f), (g), (h), (i), (j) show corresponding HNGD spectra, respectively, for $5$ ms segment around the epoch location represented by solid line.}}
	\label{fig_hngd}
	\end{figure*} 
    
%Due to the coupling between vocal-tract and excitation source during speech production, it is better to consider the spectrum only in glottal closed phase.
HNGD is found to have potential in deriving VTS for a very short segment of speech signal around GCI that mostly corresponds to the glottal closed phase as reported in~\cite{bayya2013spectro}. It is employed in this work to analyze different characteristics of VTS for sonorant sounds. The same process of deriving HNGD spectrum around each GCI in the speech signal, as in~\cite{bayya2013spectro} is used here: 
  
\begin{itemize}
\item The frequency response of ZFF as proposed in~\cite{murty2008epoch} can be represented by (\ref{ZFFeq}). The analogous time domain window function shown in (\ref{ZTW}) is used to emphasize the speech samples closest to each GCI location. This windowing method is referred as zero time windowing (ZTW)~\cite{bayya2013spectro}.
\begin{eqnarray}\label{ZFFeq}
|H(w)|= |1/(1-z^{-1})^2|_{z=e^{jw}}=1/2(1-cosw)\nonumber\\
 = 1/4sin^2(w/2)
\end{eqnarray}
   
\begin{align}\label{ZTW}       
         w[n] = \left\{ \begin{array}{ll}
         0 & n=0;\\
         1/(4sin^{2}(\pi n/(2N))) & n=1,2,.....N-1.\end{array} \right. 
\end{align}        
where, $N$ is the length of the window. 

\item Let $s(n)$ be the speech signal and corresponding epoch locations are extracted by using  ZFF signal as explained in~\cite{murty2008epoch}. This can be represented by a train of impulses as shown in (\ref{delta}), where $M$ is total number of epochs and $i_k$ is the estimated epoch location~\cite{sharma2014faster}.
   \begin{align} \label{delta}
   \sum_{k=1}^{M}\delta(n-i_k)\
   \end{align}

\item    Let $x_k(n)$ be the windowed signal derived by placing the window at each epoch location as shown in (\ref{windwd})
   \begin{align} \label{windwd}
   x_k(p)=s(p)\times w(n)\
   \end{align}
   where, $p=i_k,i_k+1,...i_k+N-1$ and $N$ is length of  window function ($w(n)$). 

\item Due to highly decaying nature of the window function, there is possibility of masking of formant peaks by over-smoothing and thereby loosing required evidences from formants. This effect of peaks merging or smoothing can be avoided by using Fourier transform phase spectra i.e. group-delay (GD) spectra instead of usual magnitude spectra~\cite{yegnanarayana1978formant}. The numerator of the GD function (NGD) ($g(w)$) of $x_k(n)$ is computed as in~\cite{bayya2013spectro}
  \begin{align}
 g(w)=X_{R}(w)Y_{R}(w)+X_{I}(w)Y_{I}(w)\
  \end{align}
where, $X(w)=X_{R}(w)+jX_{I}(w)$ is the discrete time Fourier transform (DTFT) of $x_k(n)$ and $Y(w)=Y_{R}(w)+jY_{I}(w)$ is the DTFT of $y_k(n)=nx_k(n)$. The subscripts ‘R’ and ‘I’ denote real and imaginary parts, respectively.
   
\item The spectral resolution is enhanced by successively differentiating NGD two times (DNGD), which shows sharp peaks at each formant location.

\item  In order to highlight these peaks further, HE  of the DNGD is computed which is called HNGD spectrum. 
   
\end{itemize}

For different categories of sound units, HNGD is found to have the potential to detect formant characteristics with accuracy for short window, as reported in~\cite{bayya2013spectro}. This motivate to exploit usefulness of HNGD spectrum in characterizing VTS to derive sonority feature.
 
\subsection{Effectiveness of HNGD spectrum for sonority detection}

	\begin{figure}[h]
	\centering
	\centerline{\epsfig{figure=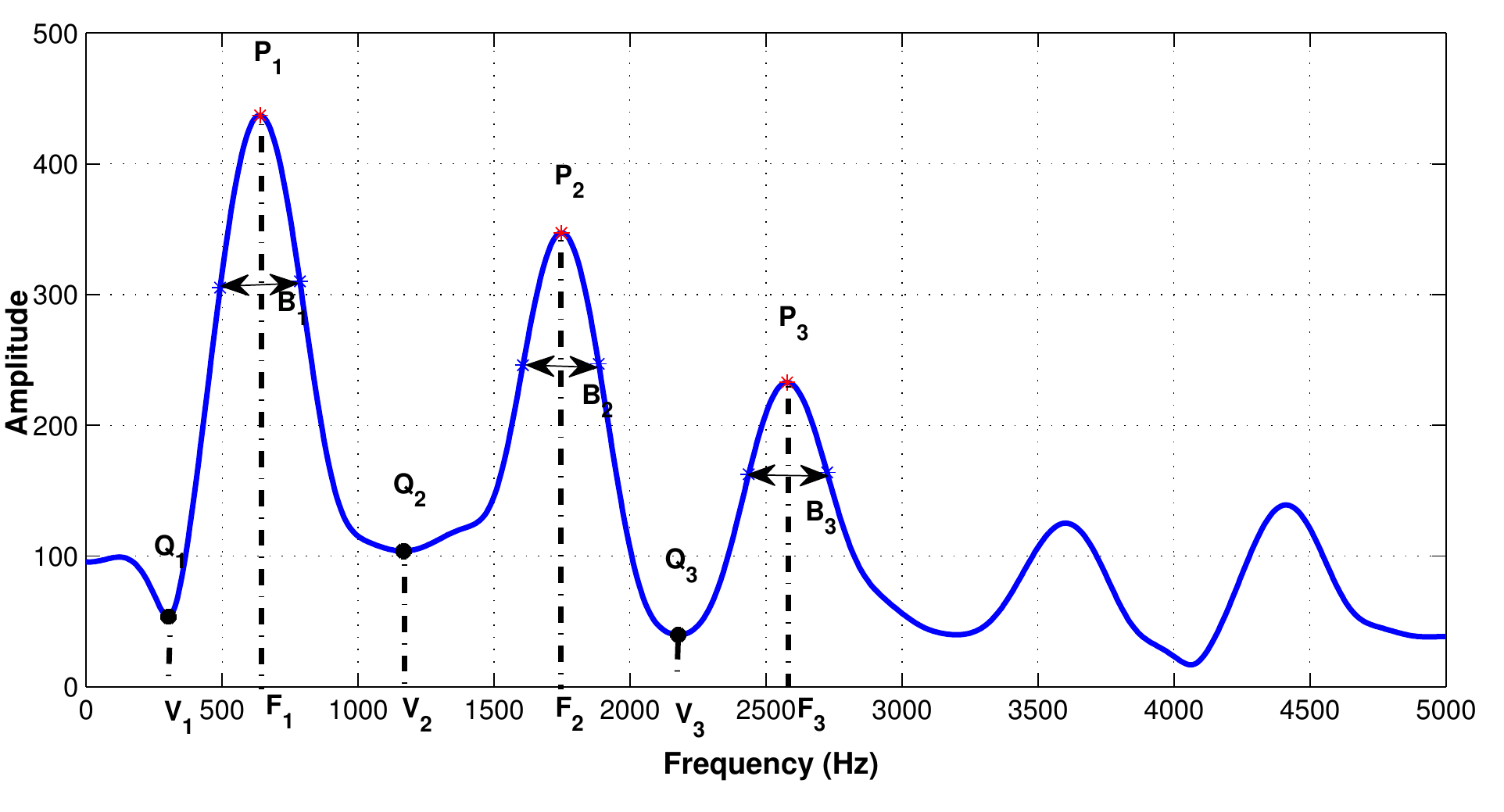,height=1.8in, width=3.5in}}
	\vspace{-.4cm}
	\caption{{\it \scriptsize Vocal-tract spectrum represented by HNGD spectrum corresponding to /eh/ showing different measurements i.e. first three formant frequency values (in Hz), amplitude of spectral peaks, frequency at spectral valleys (in Hz), amplitude of spectral valleys and bandwidth}}
	\label{fig_peak}
	\end{figure}

In order to substantiate the variation in formant structure of the HNGD spectra with respect to degree of sonority, the same is shown in Fig.~\ref{fig_hngd} for different classes of sounds. Figures~\ref{fig_hngd} (a) - (e) show $20$ ms segments of low-vowel /ah/, mid-vowel /eh/, high-vowel /ih/, semi-vowel /w/, nasal /n/, respectively. The epoch locations marked with dashed vertical lines are derived using ZFF method as described in~\cite{murty2008epoch}. Figures~\ref{fig_hngd} (f) - (j) show HNGD spectra around the epochs represented by solid lines in Fig.~\ref{fig_hngd} (a) - (e), respectively. For the spectrum of low-vowel /ah/, first three spectral peaks have higher amplitudes, higher slopes and lower bandwidths. The slope represents rate of decay of the spectrum amplitude from spectral peaks ($P_1, P_2, P_3$) to corresponding preceding spectral valleys ($Q_1, Q_2, Q_3$) as in Fig.~\ref{fig_peak}. On the other hand, mid-vowel /eh/ has lower $F_1$ and hence lower 
spectral prominence than that of the low-vowel. For high-vowel, $F_1$ decreases further. With the decrease in $F_1$ value, reduction in overall spectrum amplitude can also be observed. For semi-vowel and nasal sounds, differences between different attributes of spectra are depicted in Fig~\ref{fig_hngd}(i) and (j). Influenced by these observations, sonority feature is represented using different statistics from the HNGD spectrum.

\subsection{Proposed features of vocal-tract system to find degree of sonority } \label{Proposed feature}

        \begin{figure*}[th]
	\centering
	\centerline{\epsfig{figure=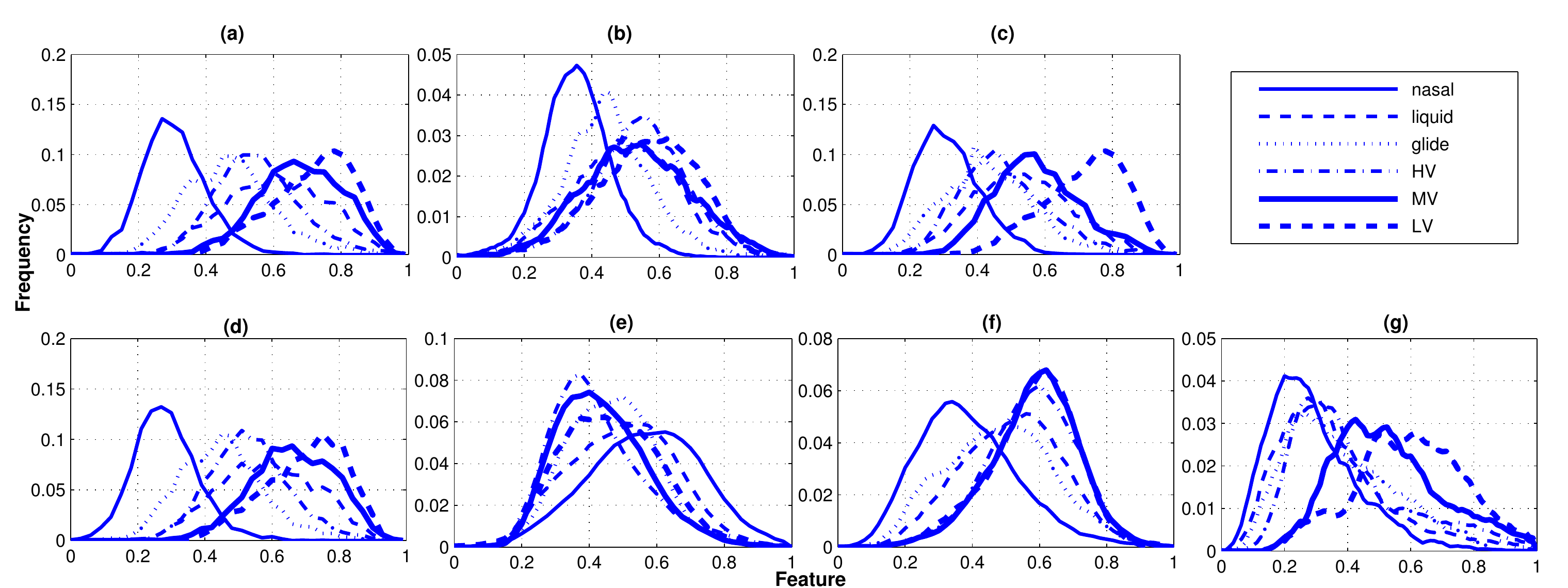,height=2.3in, width=6.3in}}
	\vspace{-.3cm}
	\caption{{\it \scriptsize Distributions of the proposed sonority features for different sonorant sound units. Distribution for feature (a)$f_1$, 
	(b)$f_2$, (c)$f_3$, (d)$f_4$, (e)$f_5$}, {\it \scriptsize (f) feature of excitation source ($f_6$)} {\it \scriptsize and (g) suprasegmental feature ($f_7$)}}
	\label{All_sys_feat2}
	\end{figure*}

 In order to find the degree of sonority associated with a sound unit, different attributes of VTS are derived from the HNGD spectrum, obtained around each epoch location. Different classes of sonorant sounds from TIMIT database used in this study are {\it nasals} ([m], [n], [ng]), {\it liquids} ([r], [l]), {\it glides} ([w], [y]), {\it high-vowels} ([ih], [iy], [uh], [uy]), {\it mid-vowels} ([eh], [ey], [oy], [ow]) and {\it low-vowels} ([aa], [ah], [ae]). These categories of sound units are segmented according to the information in TIMIT label files, succeeded by normalization with respect to the maximum value of each sound unit, for which epoch locations are derived. HNGD spectrum of energy normalized speech segment after each epoch location, is obtained as described in Section~\ref{Zero Time Windowing} which has potential to correctly characterize VTS~\cite{bayya2013spectro}. The first three formant frequencies and associated measures are of crucial importance in many speech processing studies. Therefore, 
the 
same in HNGD spectra are employed for the task of extraction of features having capability to represent sonority. The effectiveness of each of the proposed features can be justified from the distribution curves obtained for the entire TIMIT test database for different classes as shown in Figure~\ref{All_sys_feat2}. 
 
 Following measures are extracted from the estimated VTS for measuring sonority.
 
\subsubsection{Formant peak values}
The first three formant frequency values (in Hz) obtained from HNGD spectrum are $F_1$, $F_2$, $F_3$ and the corresponding amplitude of spectral peaks are
represented by $P_1$, $P_2$, $P_3$ as shown in Fig.~\ref{fig_peak}. With the increase in degree of sonority, $F_1$ (in Hz) increases. This is also reflected in the amplitude of spectral peaks, as increase in $F_1$ results in overall increase in the spectrum amplitude. The mean amplitude of first three spectral peaks is calculated, which is represented as $f_1$, where, $f_1=\frac{1}{3}\sum_{i=1}^{3} P_i$. The estimated distribution of normalized value of $f_1$ for different classes of sonorant sounds is shown in Fig.~\ref{All_sys_feat2}(a). It can be observed from Fig.~\ref{All_sys_feat2}(a) that $f_1$ may not discriminate well between different sonorant classes, but it does provide some evidence along the lines of sonority hierarchy.
   
 \subsubsection{Formant peak deviation}
When two or more formant frequencies come close together, there is an increase in spectrum value in the vicinity of these formant frequencies. The next measure for sonority measurement from VTS is the mean of relative deviation between amplitude of first three spectral peaks. Here $D_1$ and $D_2$ are differences between amplitudes of first and second  spectral peaks, and second and third spectral peaks, respectively. The mean of these differences is represented as $f_2=\frac{1}{2}\sum_{i=1}^{2}D_i$. The distribution corresponding to normalized value of $f_2$ for different sonorant classes derived from whole TIMIT test database is shown in Fig.~\ref{All_sys_feat2}(b). $f_2$ may provide some information along the sonority hierarchy.

 \subsubsection{Spectral valleys preceding the first three formant peaks}
    Along with spectral peaks, spectral valleys are also of importance for overall study of the spectrum shape. 
    Spectral valleys ($V_1, V_2, V_3$) preceding to the first three formant frequencies ($F_1, F_2, F_3$) are detected and the mean value of corresponding spectral amplitudes $Q_1, Q_2, Q_3$ is calculated. It is represented as $f_3=\frac{1}{3}\sum_{i=1}^{3} Q_i$. The distribution of normalized $f_3$ derived from segments of different sonorant classes from entire TIMIT test database is shown in Fig.~\ref{All_sys_feat2}(c).

  \subsubsection{Slope associated with each formant peak}
  In order to detect spectral prominence, slope associated with each spectral peak is also measured. To measure the slope,
  first three spectral peaks ($P_1, P_2, P_3$) corresponding to formant frequency values $F_1, F_2, F_3$ are detected. Similarly, preceding amplitude of spectral valleys ($Q_1, Q_2, Q_3$) corresponding to frequency values $V_1, V_2, V_3$ are determined as shown in Fig.~\ref{fig_peak}. Then, slope associated with each of the first three spectral peaks is calculated as follows:
  \begin{align}
  SP_1=\frac{P_1-Q_1}{F_1-V_1}; 
  SP_2=\frac{P_2-Q_2}{F_2-V_2};
  SP_3=\frac{P_3-Q_3}{F_3-V_3}\;
  \end{align}
  
   To represent this feature, average value of $SP_1, SP_2$ and $SP_3$ is calculated as, $f_4=\frac{1}{3}\sum_{i=1}^{3} SP_i$. The distributions are obtained for normalized value of $f_4$ for different sonorant classes in the TIMIT test database as shown in Fig.~\ref{All_sys_feat2}(d).
  
  \subsubsection{Formant Bandwidth}
  Formant bandwidth is directly proportional to the loss associated with vocal-tract. This may arise from different sources such as vocal-tract walls,
  viscosity, heat conduction and radiation. Hence, with more constricted vocal-tract configuration, bandwidth associated with peaks also increases. This results in decrease in degree of sonority. Before calculating the bandwidth, the spectrum is converted to log scale ($10 \log (g(w)_{hngd}))$), where, $g(w)_{hngd}$ represents HNGD spectrum. For each of the first three spectral peaks ($P_1, P_2, P_3$), corresponding $3$ dB bandwidths ($B_1, B_2, B_3$) are measured and average bandwidth is calculated ($f_5=\frac{1}{3}\sum_{i=1}^{3} B_i$). The distributions corresponding to normalized bandwidth is shown in Fig.~\ref{All_sys_feat2}(e), which decreases with the increase in sonority. 
  
   The values of each of the features $f_1, f_2, f_3, f_4, f_5$ obtained from all the frames across all instances of the six types of sounds are normalized as follows:
   
   \begin{align}
   f_i=\frac{f_i-min(f_i)}{max(f_i)-min(f_i)}\
   \end{align}
   
   where, $i$ ranges from $1$ to $5$. $min(f_i)$ and $max(f_i)$ represent minimum and maximum values of $f_i$ extracted over all classes of sonorant sounds for entire TIMIT test database. 
   
\subsection{Combined Vocal Tract feature to find degree of sonority} \label{combine_system}
\begin{table} [h]
\caption{\label{CCA} {\textit{Canonical correlation analysis (CCA) between different features of vocal-tract system}}}
\renewcommand{\arraystretch}{1}
%\vspace{2mm}
\centerline{
\scalebox{1.1}{
%\centerline{
\begin{tabular}{|c c|}
\hline
\multicolumn{1}{|c|}{\bf{Features}} & \multicolumn{1}{|c|}{\bf Correlation value}\\
\hline
\hline
\multicolumn{1}{|c|}{${\bf f_1}$ and ${\bf f_2}$} & \multicolumn{1}{|c|}{0.89}\\
\hline
\multicolumn{1}{|c|}{${\bf f_1}$ and ${\bf f_3}$} & \multicolumn{1}{|c|}{0.88}\\
\hline
\multicolumn{1}{|c|}{${\bf f_1}$ and ${\bf f_4}$} & \multicolumn{1}{|c|}{0.63}\\
\hline
\multicolumn{1}{|c|}{${\bf f_1}$ and ${\bf f_5}$} & \multicolumn{1}{|c|}{0.40}\\
\hline
\multicolumn{1}{|c|}{${\bf f_2}$ and ${\bf f_3}$} & \multicolumn{1}{|c|}{0.89}\\
\hline
\multicolumn{1}{|c|}{${\bf f_2}$ and ${\bf f_4}$} & \multicolumn{1}{|c|}{0.52}\\
\hline
\multicolumn{1}{|c|}{${\bf f_2}$ and ${\bf f_5}$} & \multicolumn{1}{|c|}{0.38}\\
\hline
\multicolumn{1}{|c|}{${\bf f_3}$ and ${\bf f_4}$} & \multicolumn{1}{|c|}{0.59}\\
\hline
\multicolumn{1}{|c|}{${\bf f_3}$ and ${\bf f_5}$} & \multicolumn{1}{|c|}{0.39}\\
\hline
\multicolumn{1}{|c|}{${\bf f_4}$ and ${\bf f_5}$} & \multicolumn{1}{|c|}{0.33}\\
\hline
\end{tabular}}
}
\end{table} 

Each of the features $f_1, f_2, f_3, f_4$ and $f_5$ are normalized and approximated by Gaussian probability density function as shown in Fig.~\ref{All_sys_feat2} (a), (b), (c), (d), (e), respectively. The distributions do not provide clear discrimination among different classes of sonorants. However, still the increasing trend of the features $f_1, f_2, f_3$ and $f_4$ from nasals to low-vowels can be observed, while $f_5$ exhibits a decreasing trend for the same. Also, some disparity in terms of overlap of distributions among different classes of sounds for each of the  features of VTS can be interpreted from Fig.~\ref{All_sys_feat2} (a)-(e). For example, in the distribution of $f_2$, a distinct overlap between the low-vowel, mid-vowel and high-vowel can be observed. $f_1$ shows less overlap between the three vowel categories along the line of sonority hierarchy. $f_2$ has lower amount of overlap between the distributions of glides and nasals. 

It can be inferred from Fig.~\ref{All_sys_feat2}(c) that, $f_3$ possess better adequacy to bring out the differences between low-vowel and mid-vowel compared to other features. In each of $f_1$, $f_3$ and $f_4$, the liquids have higher values than that of glides, whereas according to the sonority hierarchy, glides are more sonorous than the liquids. In Fig.~\ref{All_sys_feat2}(e), $f_5$ shows a correct reverse trend of feature values with respect to the sonority hierarchy. However, the extent of overlap between different classes is more compared to other features. Based on this interpretation, it can be inferred that the five derived features of vocal-tract system may carry different information. 

The redundancy among the five attributes derived from the VTS is elucidated using canonical correlation analysis (CCA)~\cite{seber2009multivariate, schott2011principles}. The correlation values derived from CCA among different pairs of features are shown in Table~\ref{CCA}. Although correlation exists between the five features of vocal-tract system, there is some extra information captured by each feature, as the correlation value is less than 1 in each case. Based on these observations, a five-dimensional feature vector of vocal-tract system is proposed in this work, which has the ability to quantify the sonority hierarchy.

\section{Excitation source information for sonority detection} \label{source_feature}

         \begin{figure}[h]
	\centering
	\centerline{\epsfig{figure=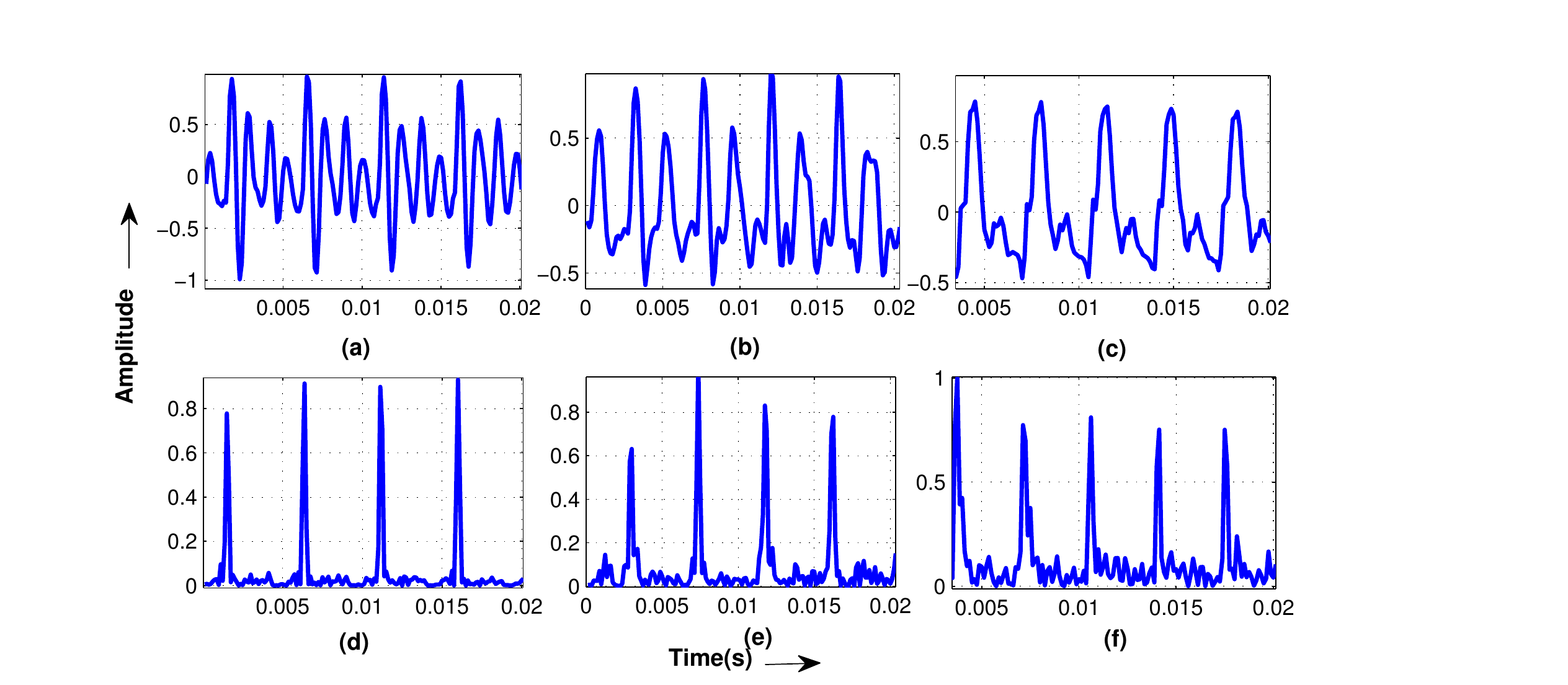,height=2in, width=3.5in}}
	\vspace{-.3cm}
	\caption{{\it Illustration of difference in nature of excitation source in vowels, semi-vowels and nasals. (a)-(c) show $20$ ms speech segment of  vowels, semi-vowels and nasals. (d)-(f) show corresponding HE of LP residual, respectively.}}
	\label{henv_speech}
	\end{figure}

         \begin{figure}[h]
	\centering
	\centerline{\epsfig{figure=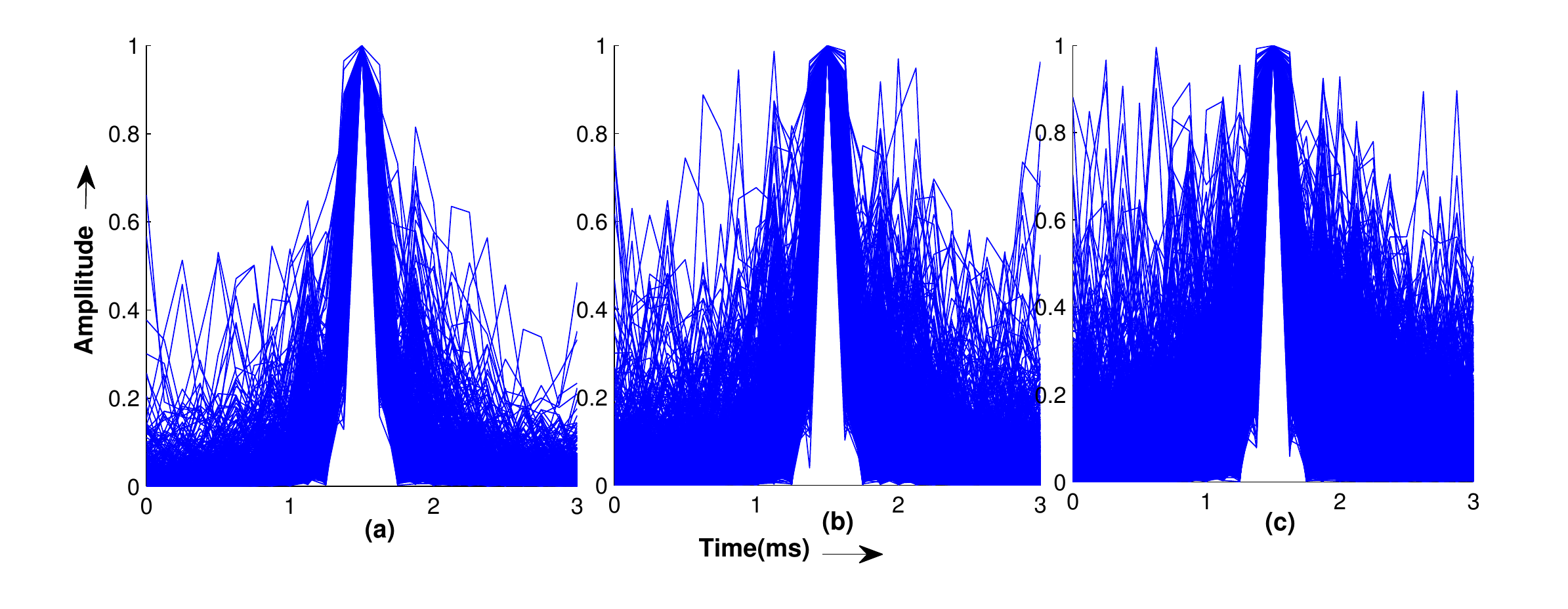,height=1.5in, width=3.3in}}
	\vspace{-.5cm}
	\caption{{\it $3$ ms duration of superimposed segments of HE of LP residual in the vicinity of impulse-like excitations for (a) Vowels, (b) semi-vowels, (c) nasals.}}
	\label{superimposed}
	\end{figure}

        \begin{figure*}[th]
	\centering
	\centerline{\epsfig{figure=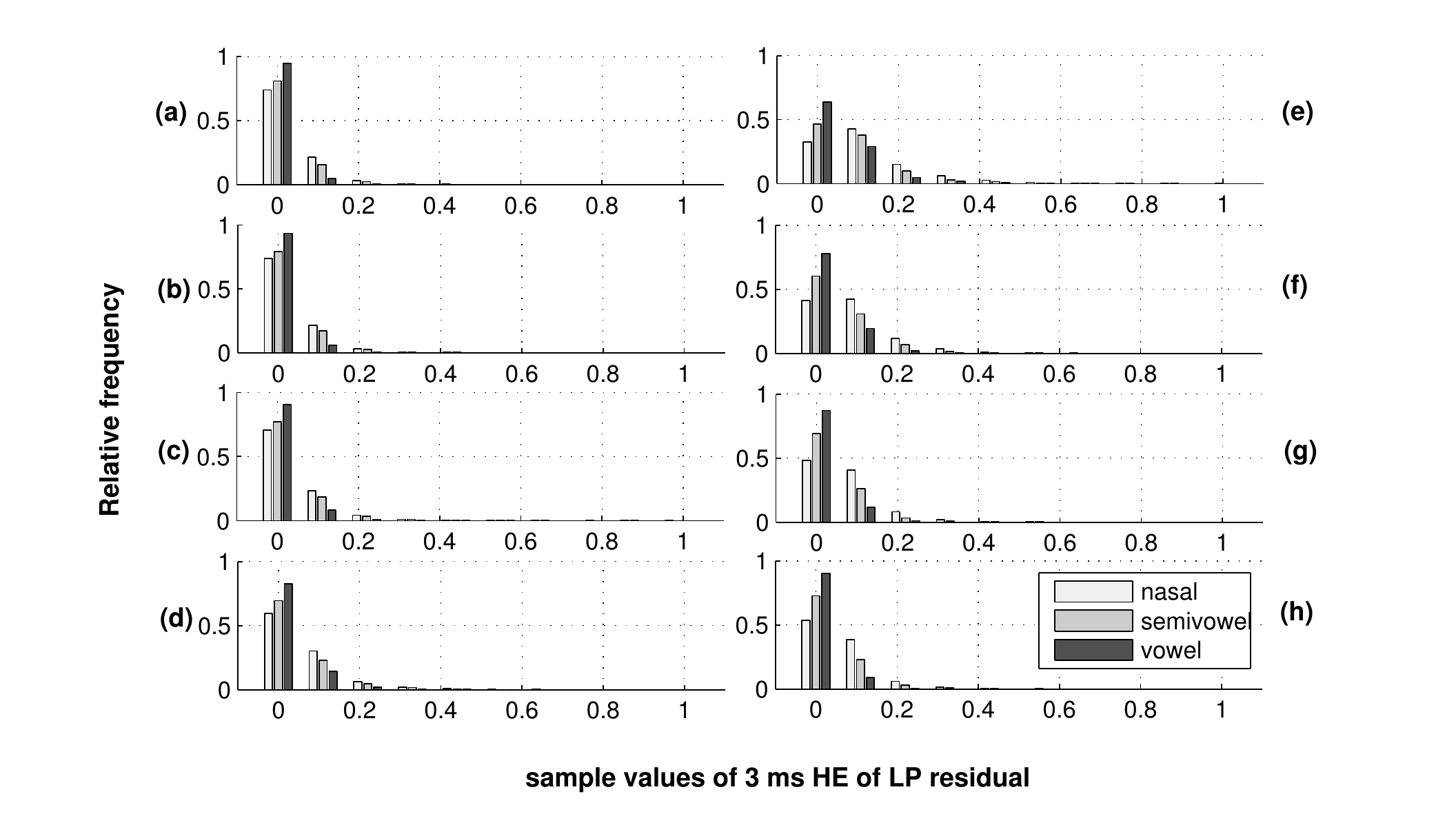,height=3in, width=6in}}
	\vspace{-.3cm}
	\caption{{\it \scriptsize Histogram plot of sample values of $3$ ms HE of LP residual. $3$ ms segment is divided into $0.25$ ms frames. (a),
	(b),(c),(d) correspond to $0$ to $1$ ms and (e), (f), (g), (h) corresponds to $2$ to $3$ ms of the $3$ ms segment}}
	\label{discrete_histogram}
	\end{figure*}
	
Strength of excitation (SoE) is related to the abruptness of glottal closure, which is maximum for an ideal impulse and corresponds to strength of differenced electro-glotto-graph (DEGG) signal at GCIs. In order to visualize how SoE changes with degree of sonority, an effective representation of SoE derived from excitation source needs to be explored. Given the speech segment of particular sound unit (vowels, semi-vowels or nasals), LP analysis can be performed to derive the LP coefficients. The residual signal is obtained by inverse filtering the speech signal using LP coefficients. The inverse filtering suppresses the vocal-tract characteristics from the speech signal and mostly contains information about the excitation source. The residual signal shows noise like characteristics in unvoiced regions and large discontinuity in voiced regions of the speech signal. This is a good approximation of excitation source signal when LP order is properly chosen~\cite{makhoul1975linear}. In this work, the LP residual 
is 
derived by performing LP analysis on overlapped segments of speech signal (size of frame =$25$ ms, frame shift = $5$ ms, LP order = 10 and sampling frequency = 8 kHz). The GCIs are manifested as large amplitude fluctuations, either in positive or negative polarity in the LP residual. This difficulty can be overcome by using the HE of LP residual~\cite{ananthapadmanabha1979epoch}. The HE $h_{e}(n)$ of LP residual $e(n)$ is defined as
  \begin{align}
   h_{e}(n)=\sqrt{e^{2}(n)+e_{h}^2(n)}\
  \end{align}
  where, $e_{h}(n)$ is Hilbert transform of $e(n)$ and in given by
  \begin{align}
   e_{h}(n)=IDFT[E_{h}(k)]
  \end{align}
where,
 \begin{align}
  	  E_{h}[k] = \left\{ \begin{array}{ll}
         -jE(k) & k=0,1,...(\frac{N}{2})-1;\\
        jE(k) & k=(\frac{N}{2}),(\frac{N}{2})+1,.....(N-1) \end{array} \right. 
 \end{align}    
 IDFT denotes inverse discrete Fourier transform and $E(k)$ is discrete Fourier transform (DFT) of $e(n)$ and $N$ is the number of points for computing DFT. 

Speech segments of $20$ ms and corresponding HE for vowel, semi-vowel and nasal are shown in Fig.~\ref{henv_speech} (a) - (c) and (d) - (f), respectively. It can be observed that, the pattern of side-lobes of each peak in HE (corresponding to GCI) is different for nasals, semi-vowels and vowels. The side-lobes have higher values with respect to peak values in case of nasals than semi-vowels. In case of vowels, the amplitude of side-lobes are further reduced than that of semi-vowels.

  	\begin{figure}[h]
	\centering
	\centerline{\epsfig{figure=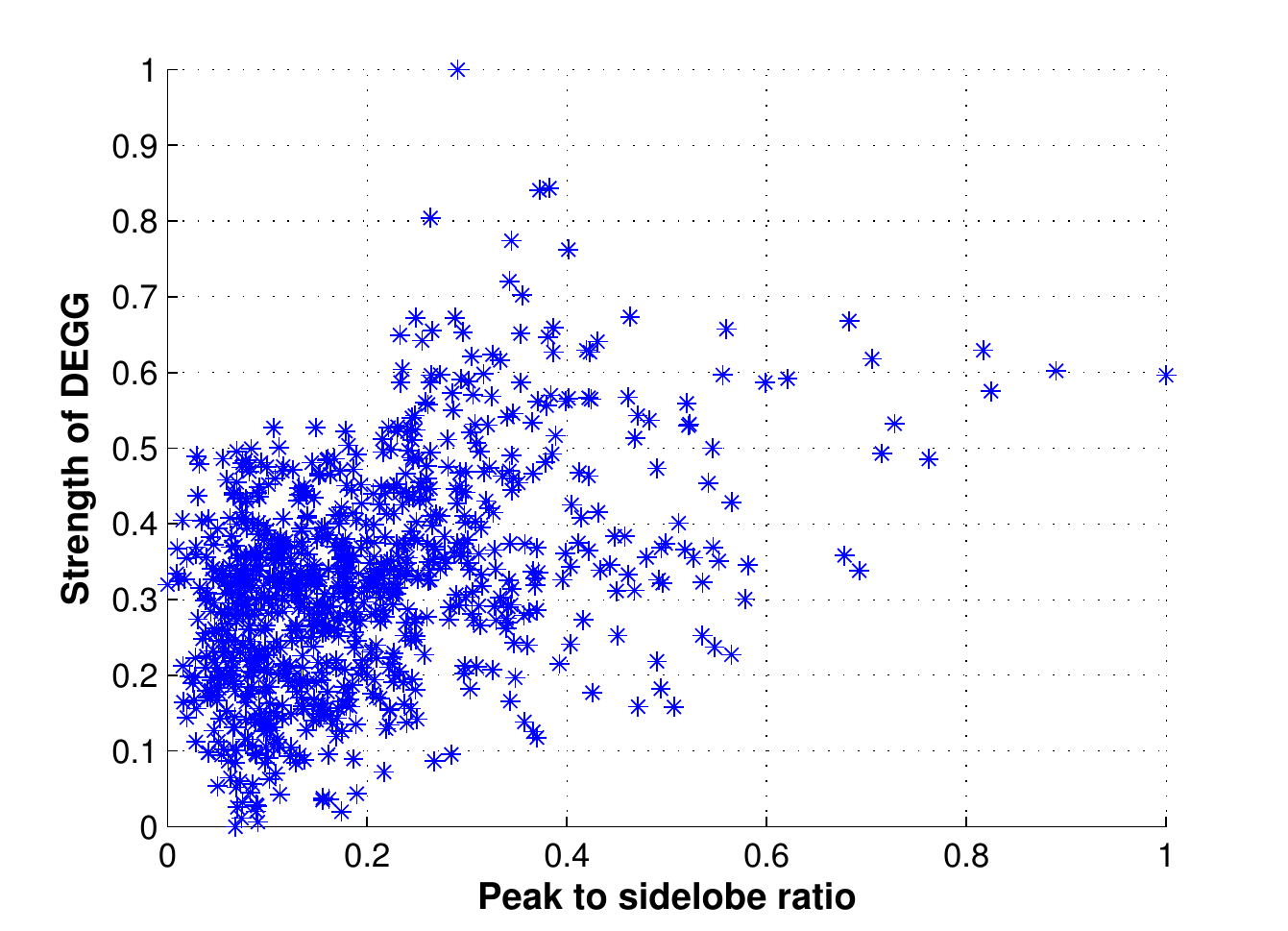,height=2in, width=3in}}
	\vspace{-.3cm}
	\caption{{\it Scatter plot of DEGG versus peak to side-lobe ratio of short segment of HE pf LP residual in the vicinity of GCIs}}
	\label{scatter_SOE}
	\end{figure}
   
 For the entire TIMIT test database, HE of LP residual of vowels, semi-vowels and nasals are obtained. The GCIs are derived from the ZFF signal and then by searching for the nearest peaks in the HE of LP residual~\cite{murty2008epoch,sharma2015improvement,sharma2016speech}. For each GCI, $1.5$ ms segment towards right and $1.5$ ms segment towards left is selected from the HE of LP residual of speech signal. These $3$ ms segments are normalized (each sample is divided by maximum value among the $3$ ms samples) and superimposed for each class (vowels, semi-vowels and nasals). The number of such superimposed frames used is equal for each class. The resulting plot is shown in Fig.~\ref{superimposed}. It can be clearly observed that the distribution of side-lobes around the center peak is different for the three classes of speech sounds. 
 
To investigate the difference among the three, the $3$ ms segment is divided further into frames of $0.25$ ms. The distribution of values for each $0.25$ ms frame is plotted using a discrete histogram as shown in Fig.~\ref{discrete_histogram}, where, (a), (b), (c), (d) correspond to first $0$ to $1$ ms (4 frames each of $0.25$ ms) and (e), (f), (g), (h) correspond to  $2$ to $3$ ms of $3$ ms of HE segment. It can be observed from Fig.~\ref{discrete_histogram} that (e), (f), (g), (h) show more discrimination between the classes (vowels, semi-vowels and nasals) than first $1$ ms frames i.e. (a), (b), (c), (d). For example: the bins corresponding to vowels, semivowels and nasals are more separated in (f) compared to that in (b). Based on this analysis, we considered only the region from $2$ to $3$ ms of the $3$ ms HE segment to quantify the source evidence. Since the distribution of values of HE of LP residual in glottal closure region is different for broad classes of sonorant sounds (vowels, semi-vowels and 
nasals), it may be appropriate to analyze the same to quantify the sonority hierarchy. 

The source feature for sonority is defined as $f_6=\frac{P}{\mu}$, where, $P$ is the value of central peak at the GCI location and $\mu$ is the mean of sample values from $2$ to $3$ ms duration in the $3$ ms HE segment. This can be referred as {\it peak to side-lobe ratio} around the epoch locations which can represent SoE. As shown in Fig.~\ref{scatter_SOE}, the SoE derived from HE of LP residual (peak to side-lobe ratio) has approximately linear correspondence with strength of DEGG signal. The distribution of peak to side-lobe ratio representing SoE for different classes of sound shows an increasing trend with the increase in sonority which can be observed from Fig.~\ref{All_sys_feat2}(f). The feature of excitation source shows a significant overlap within the vowel categories, whereas it has potential to correctly discriminate source aspect of nasals and vowels. Semi-vowels (glides and liquids) also seem to have overlapped distributions. However, the distributions of $f_6$ for each class shows less 
variance compared to that of features of vocal-tract system.
\vspace{-0.2cm}
\section{Suprasegmental Evidence for Sonority Measurement} \label{suprasegmental_feature}
\vspace{-0.3cm}
        \begin{figure}[h]
	\centering
	\centerline{\epsfig{figure=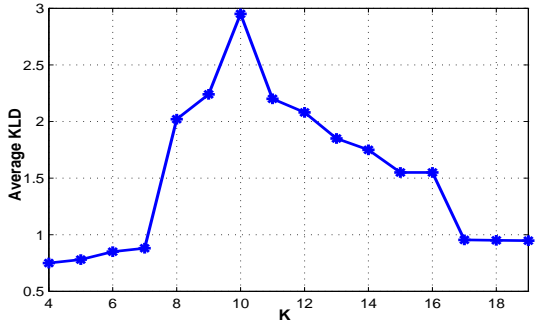,height=1.8in, width=2.8in}}
	\vspace{-.3cm}
	\caption{{\it Change in average KLD between Gaussian distributions derived from suprasegmental feature of six classes of sonorant sound with respect to the value of K.}}
	\label{Supra_KLD}
	\end{figure}

Sonorant sounds are prolonged with higher periodicity, where similar signal structure repeats for longer duration due to the slow change in vocal-tract configuration during production. This behavior of sonorant sounds can be captured by measuring similarity of speech signal samples over several pitch periods rather than just one pitch period. In this work, a suprasegmental feature is derived by computing correlation of speech signal over $K$ pitch periods as a manifestation of regularity in the structure of speech signal. If there are $M$ number of epochs in the given speech signal, $x_1, x_2,.....x_{M-1}$ are the segments corresponding to $M-1$ number of cycles starting from one epoch to the next. The similarity over $K$ number of cycles (pitch periods) is measured as follows:
\begin{align}
 f_7(i)= \frac{1}{K} \sum_{j=i+1}^{i+K} \frac{<x_i,x_j>}{\sum^{N_i} x_i^2 \sum^{N_j} x_j^2}; i=1,2,....M-1-K
\end{align}
where, $f_7(i)$ is the correlation coefficient representing suprasegmental evidence of sonorants. $<x_i, x_j>$ represents the inner product between samples corresponding to $x_i$ and $x_j$, which are $i^{th}$ and $j^{th}$ pitch cycles in the speech segment. Zero padding is performed to match the dimension of $x_i$ and $x_j$. $N_i$ and $N_j$ are the number of samples present in $i^{th}$ and $j^{th}$ cycles. $M$ is the total number of GCIs in the given speech segment and $K$ is the number of cycles over which the similarity measure is calculated. 

For finding appropriate value of $K$, the suprasegmental feature is derived by varying $K$ value from $4$ to $19$. For each value of $K$, Gaussian distributions of the six classes are obtained and average KLD measure among the six classes is calculated. The $K$ value which gives maximum KLD distance between the distribution of six sonorant classes is selected. Figure~\ref{Supra_KLD} shows that for $K=10$, the KLD distance has highest value. If the length of the speech segment is less than $10$ pitch periods, the $K$ value is changed to two less than the number of pitch periods in the signal. For $M$ number of GCIs in the speech signal, suprasegmental feature $f_7$ will have $M-1-K$ number of values. This corresponds to first $M-1-K$ number of epochs. For last $K+1$ number of epochs, the last value of feature is repeated to match the suprasegmental feature dimension with that of vocal-tract system and excitation source feature. The derived correlation feature is obtained for different categories of sonorants 
from 
TIMIT test database and the corresponding distribution is depicted in Fig.~\ref{All_sys_feat2}(g). As hypothesized, proposed suprasegmental aspect of speech signal has the adequacy to delineate the sonority hierarchy. Regardless of the significant overlap between distributions of liquids, glides and high-vowels in Fig.~\ref{All_sys_feat2}(g), it shows an increase in feature value as one moves from nasals (least sonorous) to low-vowels (most sonorous).

\section{Combination of source, system and suprasegmental evidence} \label{combination}
\begin{table*} [th]
\renewcommand{\arraystretch}{1.1}
\caption{\label{MeanVar} {\textit{Means and standard deviations (std) of different features of vocal-tract system ($f_1, f_2, f_3, f_4, f_5$), feature of excitation source ($f_6$) and suprasegmental feature ($f_7$) for different classes of sonorants (low-vowel, mid-vowel, high-vowel, liquid, glide and nasal).}}}
\vspace{-.2cm}
\centerline{
\scalebox{1}{
%\centerline{
\begin{tabular}{|c c c c c c c c c c c c c|}
\hline
\multicolumn{1}{|c|}{\bf{Evidence}} & \multicolumn{2}{|c|}{\bf{Low-vowel}} & \multicolumn{2}{|c|}{\bf{Mid-vowel}} &  \multicolumn{2}{|c|}{\bf{High-vowel}} & \multicolumn{2}{|c|}{\bf{Glide}} & \multicolumn{2}{|c|}{\bf{Liquid}} & \multicolumn{2}{|c|}{\bf{Nasal}} \\
\cline{2-13}
\multicolumn{1}{|c|}{\bf{}} &  \multicolumn{1}{|c|}{mean} &  \multicolumn{1}{|c|}{std} &  \multicolumn{1}{|c|}{mean} &  \multicolumn{1}{|c|}{std} &  \multicolumn{1}{|c|}{mean} &  \multicolumn{1}{|c|}{std} &  \multicolumn{1}{|c|}{mean} &  \multicolumn{1}{|c|}{std} &  \multicolumn{1}{|c|}{mean} &  \multicolumn{1}{|c|}{std} &  \multicolumn{1}{|c|}{mean} &  \multicolumn{1}{|c|}{std}\\
\hline
\hline
\multicolumn{1}{|c|}{\bf{Formant Peak Values ($f_1$)}} &  \multicolumn{1}{|c|}{0.73} &  \multicolumn{1}{|c|}{0.11} &  \multicolumn{1}{|c|}{0.69} &  \multicolumn{1}{|c|}{0.12} &  \multicolumn{1}{|c|}{0.56} &  \multicolumn{1}{|c|}{0.12} &  \multicolumn{1}{|c|}{0.48} &  \multicolumn{1}{|c|}{0.13} &  \multicolumn{1}{|c|}{0.62} &  \multicolumn{1}{|c|}{0.14} &  \multicolumn{1}{|c|}{0.32} &  \multicolumn{1}{|c|}{0.09}\\
\hline
\multicolumn{1}{|c|}{\bf{Formant peak deviation ($f_2$)}} &  \multicolumn{1}{|c|}{0.60} &  \multicolumn{1}{|c|}{0.14} &  \multicolumn{1}{|c|}{0.56} &  \multicolumn{1}{|c|}{0.14} &  \multicolumn{1}{|c|}{0.54} &  \multicolumn{1}{|c|}{0.12} &  \multicolumn{1}{|c|}{0.46} &  \multicolumn{1}{|c|}{0.11} &  \multicolumn{1}{|c|}{0.53} &  \multicolumn{1}{|c|}{0.14} &  \multicolumn{1}{|c|}{0.38} &  \multicolumn{1}{|c|}{0.08}\\
\hline
\multicolumn{1}{|c|}{\bf{Spectral valleys ($f_3$)}} &  \multicolumn{1}{|c|}{0.62} &  \multicolumn{1}{|c|}{0.12} &  \multicolumn{1}{|c|}{0.59} &  \multicolumn{1}{|c|}{0.12} &  \multicolumn{1}{|c|}{0.49} &  \multicolumn{1}{|c|}{0.13} &  \multicolumn{1}{|c|}{0.45} &  \multicolumn{1}{|c|}{0.13} &  \multicolumn{1}{|c|}{0.55} &  \multicolumn{1}{|c|}{0.14} &  \multicolumn{1}{|c|}{0.33} &  \multicolumn{1}{|c|}{0.09}\\
\hline
\multicolumn{1}{|c|}{\bf{Slope ($f_4$)}} &  \multicolumn{1}{|c|}{0.71} &  \multicolumn{1}{|c|}{0.12} &  \multicolumn{1}{|c|}{0.67} &  \multicolumn{1}{|c|}{0.12} &  \multicolumn{1}{|c|}{0.54} &  \multicolumn{1}{|c|}{0.11} &  \multicolumn{1}{|c|}{0.46} &  \multicolumn{1}{|c|}{0.12} &  \multicolumn{1}{|c|}{0.60} &  \multicolumn{1}{|c|}{0.14} &  \multicolumn{1}{|c|}{0.29} &  \multicolumn{1}{|c|}{0.09}\\
\hline
\multicolumn{1}{|c|}{\bf{Formant Bandwidth ($f_5$)}} &  \multicolumn{1}{|c|}{0.55} &  \multicolumn{1}{|c|}{0.05} &  \multicolumn{1}{|c|}{0.58} &  \multicolumn{1}{|c|}{0.05} &  \multicolumn{1}{|c|}{0.57} &  \multicolumn{1}{|c|}{0.05} &  \multicolumn{1}{|c|}{0.59} &  \multicolumn{1}{|c|}{0.05} &  \multicolumn{1}{|c|}{0.61} &  \multicolumn{1}{|c|}{0.06} &  \multicolumn{1}{|c|}{0.63} &  \multicolumn{1}{|c|}{0.06}\\
\hline
\multicolumn{1}{|c|}{\bf{Source($f_6$)}} &  \multicolumn{1}{|c|}{0.29} &  \multicolumn{1}{|c|}{0.06} &  \multicolumn{1}{|c|}{0.29} &  \multicolumn{1}{|c|}{0.06} &  \multicolumn{1}{|c|}{0.29} &  \multicolumn{1}{|c|}{0.06} &  \multicolumn{1}{|c|}{0.24} &  \multicolumn{1}{|c|}{0.08} &  \multicolumn{1}{|c|}{0.27} &  \multicolumn{1}{|c|}{0.08} &  \multicolumn{1}{|c|}{0.20} &  \multicolumn{1}{|c|}{0.08}\\
\hline
\multicolumn{1}{|c|}{\bf{Suprasegmental($f_7$)}} &  \multicolumn{1}{|c|}{0.49} &  \multicolumn{1}{|c|}{0.14} &  \multicolumn{1}{|c|}{0.44} &  \multicolumn{1}{|c|}{0.15} &  \multicolumn{1}{|c|}{0.34} &  \multicolumn{1}{|c|}{0.16} &  \multicolumn{1}{|c|}{0.32} &  \multicolumn{1}{|c|}{0.15} &  \multicolumn{1}{|c|}{0.29} &  \multicolumn{1}{|c|}{0.14} &  \multicolumn{1}{|c|}{0.24} &  \multicolumn{1}{|c|}{0.11}\\
\hline
\end{tabular}}
}
\end{table*}

The means and standard deviations of each of the derived features are shown in Table~\ref{MeanVar}. As elaborated in Section~\ref{combine_system}, the means and standard deviations of five different features of vocal-tract system carry different information regarding the degree of sonority associated with. As observed from Table~\ref{MeanVar}, from low-vowels to nasals, the mean values of $f_1, f_2, f_3$ and $f_4$ decrease sequentially with a disparity in case of glides and liquids. The latter having higher mean value than the former in case of all the four features. It can be observed that the mean values of $f_5$ increase from low-vowels to nasals. The deviation in mean values of $f_5$ among different classes is less. Also, the standard deviation values of $f_5$ are low compared to other features of vocal tract system.

From production point of view, the difference between glides and liquids is that, in case of liquids the constriction is shorter than that of the glides. This results in higher $F_1$ for liquids than glides. Moreover, the acoustic path in the oral cavity for liquids contains side branch or parallel paths unlike glides. This introduces extra poles and zeros in the spectrum of liquids which lead to higher values of features of vocal-tract system for liquids than glides. The pattern of mean values of suprasegmental feature is found to have good correlation with the degree of sonority. All the evidences derived from three different perspectives of sonorant sounds demonstrate unique trend with the change in degree of sonority. To obtain a faithful feature representation of sonority, the combination of features of vocal-tract system, feature of excitation source and suprasegmental feature may be helpful. All the seven evidences have one value at each epoch location. 
   
For each of the seven features, six Gaussian distributions can be derived representing six classes of sonorant sounds. The distance between each pair of Gaussian probability density function can be measured by Kullback Leibler divergence (KLD)~\cite{cover2012elements} as given by (\ref{KLD}).
  
  \begin{eqnarray}\nonumber \label{KLD}
D_{KL}(A,B)=\frac{1}{2}\left\{\frac{\sigma_A^2}{\sigma_B^2}+\frac{\sigma_B^2}{\sigma_A^2}\right \}-1\\
+\frac{1}{2}\{\mu_A-\mu_B\}^2\left\{\frac{1}{\sigma_A^2}+\frac{1}{\sigma_B^2}\right \}\
\end{eqnarray}

 where, $A$ and $B$ are two univariate Gaussian distributions with mean $\mu_A$, $\mu_B$ and standard deviation $\sigma_A$, $\sigma_B$, respectively. Here $A$ and $B$ represent samples of one feature for two classes of sonorant sounds. As there are $6$ classes of sonorant sounds, each feature will have $6$ Gaussian distributions i.e. $15$ pairs of distributions as shown in Fig.~\ref{All_sys_feat2}. The average KLD distance measure is calculated for each of the seven features over these $15$ pairs of distribution as in (\ref{KLD_avg}). The average KLD distance for each feature is tabulated in Table~\ref{KLD_table}.
  
\begin{eqnarray} \label{KLD_avg}
\{D_{KL}(A,B)\}_{avg}= \frac{1}{15}\sum_{i=1}^{15} D_{KL}(A,B)_i\
\end{eqnarray}  

 The seven features shown in Table~\ref{KLD_table} have difference in terms of their ability to differentiate between the classes of sonorant sounds. High value of KLD represents greater ability of the feature to discriminate different classes of sonorants and hence more weight should be assigned to that particular feature dimension. Based on the average KLD between different classes of sound, weights corresponding to each of the seven features ($w_i$) are derived such that 
 \begin{eqnarray} \label{Weights1}   
 \sum_{i=1}^7 w_i=&1\
 \end{eqnarray}
   where,
   \begin{eqnarray} \label{Weights2} 
   w_i=&\frac{[\{D_{KL}(A,B)\}_{avg}]_{f_i}}{\sum_{i=1}^7 [\{D_{KL}(A,B)\}_{avg}]_{f_i}}\
   \end{eqnarray}
 The weights assigned to each of the seven features according to their potential to classify different sonorant sounds are also shown in Table~\ref{KLD_table}. Thus a competent representation of degree of sonority associated with a sound unit is derived in this work.
 
\begin{figure}[h]
	\centering
	\centerline{\epsfig{figure=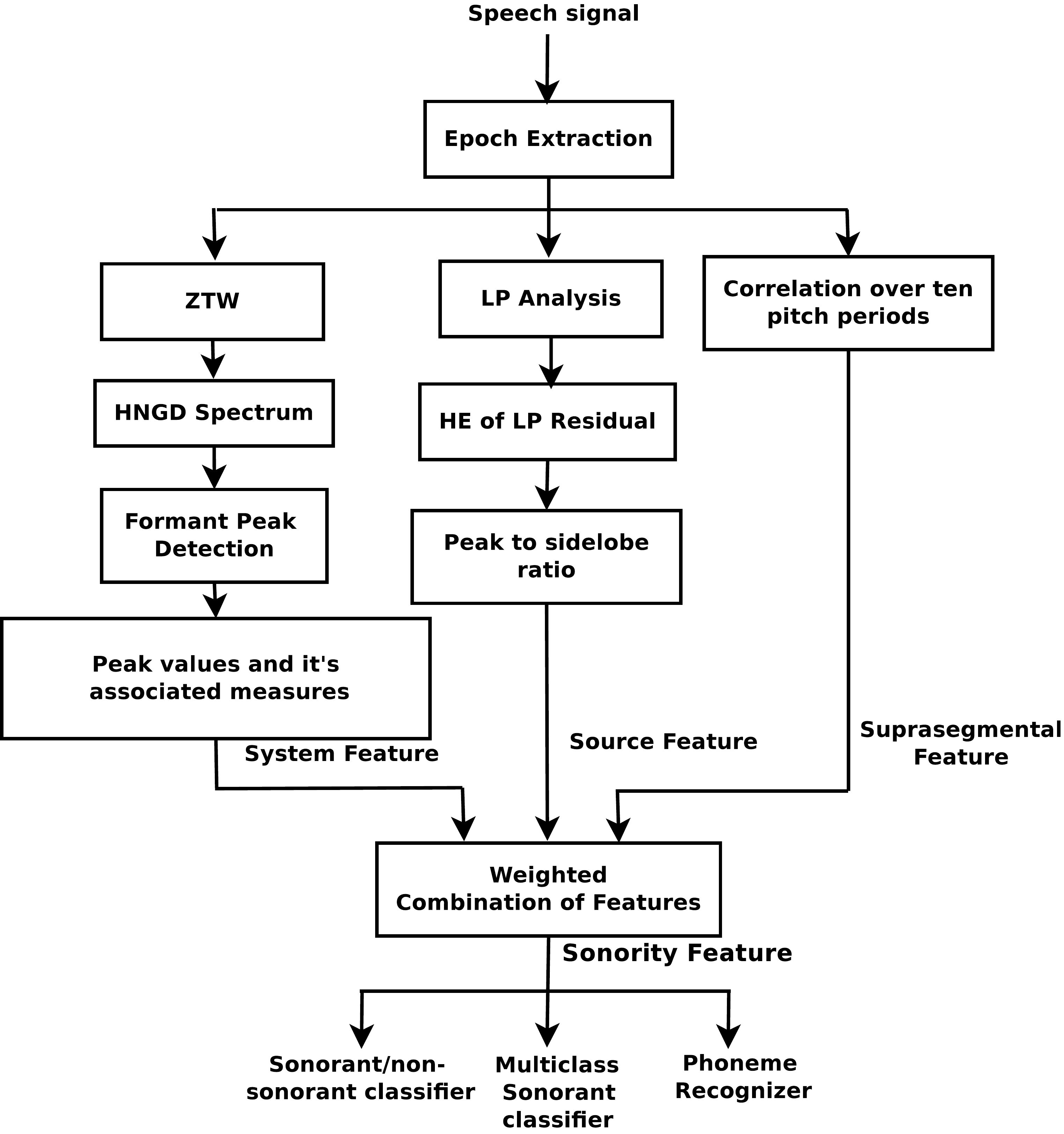,height=3.5in, width=3in}}
	\vspace{-.2cm}
	\caption{{\it Overall block diagram of the proposed sonority feature extraction from speech signal, where vocal-tract system, excitation source and suprasegmental features are derived from HNGD spectrum, HE of LP residual and speech signal, respectively. These features are combined to derive the sonority feature.}}
	\label{Block_diagram}
	\end{figure}

\begin{table} [th]
\caption{\label{KLD_table} {\textit{Average KLD between Gaussian distributions of six classes of sonorant sounds and corresponding weights  assigned for different features of vocal-tract system, exciation source and suprasegmental feature.}}}
\renewcommand{\arraystretch}{1}
%\vspace{2mm}
\centerline{
\scalebox{0.95}{
%\centerline{
\begin{tabular}{|c c c|}
\hline
\multicolumn{1}{|c|}{\bf{Features}} & \multicolumn{1}{|c|}{\bf Average KLD}  & \multicolumn{1}{|c|}{\bf Weights}\\
\hline
\hline
\multicolumn{1}{|c|}{\bf{Formant Peak Values ($f_1$)}} & \multicolumn{1}{|c|}{1.14} & \multicolumn{1}{|c|}{0.1049}\\
\hline
\multicolumn{1}{|c|}{\bf{Formant peak deviation ($f_2$)}} & \multicolumn{1}{|c|}{0.95} & \multicolumn{1}{|c|}{0.0874}\\
\hline
\multicolumn{1}{|c|}{\bf{Spectral valleys ($f_3$)}} & \multicolumn{1}{|c|}{1.10} & \multicolumn{1}{|c|}{0.1012}\\
\hline
\multicolumn{1}{|c|}{\bf{Slope ($f_4$)}} & \multicolumn{1}{|c|}{1.09} & \multicolumn{1}{|c|}{0.1003}\\
\hline
\multicolumn{1}{|c|}{\bf{Formant Bandwidth ($f_5$)}} & \multicolumn{1}{|c|}{1.62} & \multicolumn{1}{|c|}{0.1490}\\
\hline
\multicolumn{1}{|c|}{\bf{Source ($f_6$)}} & \multicolumn{1}{|c|}{2.02} & \multicolumn{1}{|c|}{ 0.1858}\\
\hline
\multicolumn{1}{|c|}{\bf{Suprasegmental ($f_7$)}} & \multicolumn{1}{|c|}{2.95} & \multicolumn{1}{|c|}{0.2714}\\
\hline
\end{tabular}}
}
\end{table}  
The overall block diagram of the proposed work is depicted in Fig.~\ref{Block_diagram}. Three different features are derived using the knowledge of vocal-tract system, excitation source and suprasegmental aspects of sonorants. To derive the feature of vocal-tract system, ZTW is performed around each epoch location of speech signal. For the windowed segments, HNGD spectra are derived. Feature of excitation source is derived from the HE of LP residual of speech signal. In contrast to these two evidences, the suprasegmental feature is derived from correlation of speech signal over ten pitch periods. The three evidences are weighted and fused together to derive the seven-dimensional sonority evidence  (vocal-tract system (five-dimension), excitation source (one-dimension) and suprasegmental feature (one-dimension)). The implementation for extraction of this sonority feature is released in the following link~\footnote{https://github.com/bidishasharma/Extract-Sonority-Feature}. The evidence is further utilized in the task of sonorant/non-sonorant classification, multiclass sonorant classification and phoneme recognition to verify the efficacy of the proposed feature.

\vspace{-0.2cm}
\section{Experimental Evaluation} \label{evaluation}

\begin{table} [th]
\caption{\label{comparision} {\textit{Comparison of performance of proposed feature (using SVM) and existing feature using hierarchical algorithm (within braces) as shown in~\cite{dumpala2015robust} in sonorant/non-sonorant segmentation on utterances from TIMIT database in both clean speech and noisy speech across different SNR levels.}}}
\renewcommand{\arraystretch}{1.5}
\vspace{-.2cm}
\centerline{
\scalebox{0.75}{
%\centerline{
\begin{tabular}{|c c c c c c c|}
\hline
\multicolumn{1}{|c|}{\bf{}} & \multicolumn{6}{|c|}{\bf Proposed feature (Existing Feature)}\\
\cline{2-7}
\multicolumn{1}{|c|}{\bf{SNR}} & \multicolumn{3}{|c|}{\bf Epoch based results} &  \multicolumn{3}{|c|}{ \bf Frame based results}\\
\cline{2-7}
\multicolumn{1}{|c|}{\bf{}} & \multicolumn{1}{|c|}{\bf Acc(\%)} &  \multicolumn{1}{|c|}{\bf TPR(\%)} &  \multicolumn{1}{|c|}{\bf FAR(\%)} &  \multicolumn{1}{|c|}{\bf Acc(\%)} &  \multicolumn{1}{|c|}{\bf TPR(\%)} &  \multicolumn{1}{|c|}{\bf FAR(\%)}\\
\hline
\hline
\multicolumn{1}{|c|}{\bf{clean}} & \multicolumn{1}{|c|}{ 96.3 (93.9)} &  \multicolumn{1}{|c|}{ 98.5 (94.5)} &  \multicolumn{1}{|c|}{ 5.5 (7.5)} &  \multicolumn{1}{|c|}{ 95.0 (92.8)} &  \multicolumn{1}{|c|}{ 97.3 (93.6)} &  \multicolumn{1}{|c|}{ 6.8 (8.0)}\\
\hline
\multicolumn{1}{|c|}{\bf{30 dB}} & \multicolumn{1}{|c|}{ 96.0 (93.9)} &  \multicolumn{1}{|c|}{ 98.4 (94.5)} &  \multicolumn{1}{|c|}{ 5.7 (7.6)} &  \multicolumn{1}{|c|}{ 94.8 (92.8)} &  \multicolumn{1}{|c|}{ 96.7 (93.6)} &  \multicolumn{1}{|c|}{ 7.6 (8.0)}\\
\hline
\multicolumn{1}{|c|}{\bf{20 dB}} & \multicolumn{1}{|c|}{ 95.4 (94.39)} &  \multicolumn{1}{|c|}{ 96.6 (94.4)} &  \multicolumn{1}{|c|}{ 6.2 (7.7)} &  \multicolumn{1}{|c|}{ 93.5 (92.7)} &  \multicolumn{1}{|c|}{ 95.8 (93.5)} &  \multicolumn{1}{|c|}{ 8.0 (8.1)}\\
\hline
\multicolumn{1}{|c|}{\bf{10 dB}} & \multicolumn{1}{|c|}{ 95.0 (93.4)} &  \multicolumn{1}{|c|}{94.3 (94.0)} &  \multicolumn{1}{|c|}{ 6.8 (8.5)} &  \multicolumn{1}{|c|}{ 93.4 (92.1)} &  \multicolumn{1}{|c|}{ 95.3 (92.9)} &  \multicolumn{1}{|c|}{ 8.4 (9.0)}\\
\hline
\multicolumn{1}{|c|}{\bf{5 dB}} & \multicolumn{1}{|c|}{ 93.4 (92.4)} &  \multicolumn{1}{|c|}{93.6 (93.1)} &  \multicolumn{1}{|c|}{ 8.3 (8.9)} &  \multicolumn{1}{|c|}{ 93.0 (91.0)} &  \multicolumn{1}{|c|}{ 92.8 (91.9)} &  \multicolumn{1}{|c|}{ 9.3 (9.5)}\\
\hline
\multicolumn{1}{|c|}{\bf{0 dB}} & \multicolumn{1}{|c|}{ 90.5 (90.7)} &  \multicolumn{1}{|c|}{ 91.4 (91.0)} &  \multicolumn{1}{|c|}{9.5 (9.9)} &  \multicolumn{1}{|c|}{ 90.0 (89.6)} &  \multicolumn{1}{|c|}{ 90.7 (89.9)} &  \multicolumn{1}{|c|}{ 10.3 (10.6)}\\
\hline
\end{tabular}}
}
\end{table}  
The distribution of the proposed sonority evidence correlates well with the sonority hierarchy as can be observed from Fig.~\ref{All_sys_feat2} and Table~\ref{MeanVar}. To establish the efficacy of the proposed seven-dimensional sonority feature vector in different speech processing applications, the following classification experiments are performed. 

\subsection{Sonorant/non-sonorant classification}
The first level of classification that exploits the usefulness of prospective features representing sonority is sonorant/non-sonorant classification. In~\cite{dumpala2015robust}, it has been demonstrated that the attributes derived from speech signal like zero frequency resonator (ZFR) signal energy, slope of ZFR signal around epoch locations and dominant resonance frequency (DRF), can be used for the task of sonorant/non-sonorant segmentation, both at frame and epoch levels. An hierarchical algorithm is used for the classification task. To compare the effectiveness of the proposed feature with the features used in~\cite{dumpala2015robust}, a sonorant/non-sonorant classifier using support-vector-machine (SVM) (with radial basis function (RBF) kernel, $c=16$, $\gamma=4$) is developed using the proposed sonority feature vector. The training and testing feature vectors are derived from all SI and SX utterances of TIMIT train and test databases, respectively. This is followed by feature normalization to make 
the 
feature values within zero to one range. Similar normalization is performed in training and testing of clean and noisy speech. The same SVM classifier trained using clean speech is employed in the testing of sentences mixed with white noise across various SNR levels. 

To demonstrate the robustness of the features for classification, the performance evaluation parameters used are: number of epochs/frames correctly detected in the sonorant regions (true positive rate (TPR)), number of spurious epochs/frames hypothesized in the non-sonorant regions (false alarm rate (FAR)) and total number of correctly detected epochs/frames in both the sonorant and non-sonorant regions (accuracy (Acc)). As shown in Table~\ref{comparision}, the proposed feature can segment sonorant regions with more accuracy compared to the existing method (within braces). Table~\ref{comparision} shows that the proposed feature has better ability to classify sonorant/non-sonorant segments from the given speech signal.

\subsection{Classification of sonorant sounds into different classes}\label{misclassification_explain}
The primary motivation of this work is to derive feature to characterize the degree of sonority associated with a sound unit. The straightforward way to validate this would be to develop a multi-class sonorant classifier. Each class represents different sonorant sounds (low-vowels, mid-vowels, high-vowels, liquids, glides and nasals). As described in Section~\ref{combination}, the proposed seven-dimensional sonority feature is derived for each class of sonorant sounds for the entire TIMIT test database. This is followed by normalization to make the feature value within the range of $0$ to $1$. Individual feature dimension consists of a single value at each epoch location. A six-class SVM classifier (with RBF kernel, $c=256$, $\gamma=16$) has been developed using the normalized sonority feature vector. Values of parameters, $c$ and $\gamma$ are set using train-test 5-fold cross validation for the entire TIMIT test database. For the optimized value of $c$ and $\gamma$, the six-class SVM 
model is trained using randomly chosen $80\%$ of TIMIT-test data. The rest $20\%$ data is used for testing. 

The classification accuracy of each class and confusion among different classes are reported in Table~\ref{six_class_epoch1}. The average accuracy achieved is $66.55\%$. The accuracy is observed to be the lowest for liquids and highest for nasals. It can be interpreted from Table~\ref{six_class_epoch1} that, $14.41\%$ of low-vowels are misclassified as mid-vowels. This is due to the fact that  the properties of low-vowels and mid-vowels are close to each other. Moreover, as observed from Fig.~\ref{All_sys_feat2}, formant bandwidth and feature of excitation source exhibit overlap between the two classes. As the height of the tongue body for mid-vowels is intermediate between that of high and low-vowels, it affects the constriction size and length. This in-turn alters the VTS evidences. 

Although the vocal-tract constriction in case of liquids is narrower than the glides resulting in wider $F_1$ bandwidth for liquids, the length of constriction is shorter in case of liquids. This increases $F_1$ for liquids and introduces confusion between glides and liquids. Thus there is possibility of confusion of liquids with low-vowels and mid-vowels. This is evident from $1^{st}$, $2^{nd}$ and $5^{th}$ rows of Table~\ref{six_class_epoch1}. The common attribute of liquids with vowels is that, in both cases air flows through the constriction without pressure drop. As a result, the vocal-folds continue to vibrate in the period of constriction. In the distribution of feature of excitation source in Fig.~\ref{All_sys_feat2}(f), confusion between glides and liquids can be apparently observed. As reported in Table~\ref{six_class_epoch1}, majority of misclassification of high-vowels is due to the confusion with mid-vowels and glides. The configuration of vocal-tract for glides may also change based on the 
preceding vowels. A glide adjacent to high-vowel is produced with more constricted structure compared to the one preceded or followed by a low-vowel. Therefore, when a glide is contiguous with low-vowel or mid-vowel, due to less constriction, $F_1$ may increase. The bandwidth may decrease compared to the glide that is adjacent with a high-vowel. 

The proposed features are analogous to formant based measures and do not use the temporal information of nearby sounds. Therefore, there is a possibility of misclassification of each category to its adjacent category of sound in the sonority hierarchy. It is notable from Fig.~\ref{All_sys_feat2} that, compared to other categories of sonorants, the distribution corresponding to nasals has less overlap with other distributions. Only in case of suprasegmental feature in Fig.~\ref{All_sys_feat2}(g), some confusion with nasals and other categories is observable. This correlates with highest accuracy for nasals as reported in Table~\ref{six_class_epoch1}. As the front part of vocal-tract is completely closed during nasal murmur, the first formant frequency and its prominence eventually decreases with a weak second formant followed by an extended valley in the VTS. This is more contrasting with other sonorants. However, the common acoustic behavior of nasals and glides is that, the vocal-fold does not change the 
vibration pattern before and after the constriction happens. Based on this discussion and the classification accuracy of sonorants presented in Table~\ref{six_class_epoch1}, it can be inferred that the proposed features have ability to quantify sonority level associated with a sound unit. Although, some aspects of the speech signal corresponding to a specific category of sound unit may vary based on the adjacent sound units present.

  \begin{table} [th]
\caption{\label{six_class_epoch1} {\textit{Classification accuracy (epoch level) of different sonorant sounds from TIMIT test database using SVM ($c=256,\gamma=16$) obtained by employing the proposed seven-dimensional sonority feature}}}
\renewcommand{\arraystretch}{1.2}
\vspace{-.1cm}
\centerline{
\scalebox{0.8}{
%\centerline{
\begin{tabular}{|c c c c c c c|}
\hline
\multicolumn{1}{|c|}{\bf Category} & \multicolumn{6}{|c|}{\bf \% Accuracy}\\
\cline{2-7}
\multicolumn{1}{|c|}{\bf } & \multicolumn{1}{|c|}{\bf Low-vowel} &  \multicolumn{1}{|c|}{ \bf Mid-vowel} &  \multicolumn{1}{|c|}{ \bf High-vowel} &  \multicolumn{1}{|c|}{ \bf Glide} &  \multicolumn{1}{|c|}{ \bf Liquid}&  \multicolumn{1}{|c|}{ \bf Nasal}\\
\hline
\hline
\multicolumn{1}{|c|}{\bf Low-vowel} & \multicolumn{1}{|c|}{\bf 68.0} &  \multicolumn{1}{|c|}{14.4} &  \multicolumn{1}{|c|}{4.1} &  \multicolumn{1}{|c|}{2.8} &  \multicolumn{1}{|c|}{9.2}&  \multicolumn{1}{|c|}{1.5}\\
\hline
\multicolumn{1}{|c|}{\bf Mid-vowel} & \multicolumn{1}{|c|}{ 9.8} &  \multicolumn{1}{|c|}{ \bf 63.9} &  \multicolumn{1}{|c|}{9.2} &  \multicolumn{1}{|c|}{4.5} &  \multicolumn{1}{|c|}{10.9}&  \multicolumn{1}{|c|}{1.7}\\
\hline
\multicolumn{1}{|c|}{\bf High-vowel} & \multicolumn{1}{|c|}{1.7} &  \multicolumn{1}{|c|}{10.3} &  \multicolumn{1}{|c|}{ \bf 67.3} &  \multicolumn{1}{|c|}{11.7} &  \multicolumn{1}{|c|}{4.6}&  \multicolumn{1}{|c|}{ 4.4}\\
\hline
\multicolumn{1}{|c|}{\bf Glide} & \multicolumn{1}{|c|}{1.4} &  \multicolumn{1}{|c|}{6.4} &  \multicolumn{1}{|c|}{12.7} &  \multicolumn{1}{|c|}{\bf 59.4} &  \multicolumn{1}{|c|}{6.7}&  \multicolumn{1}{|c|}{13.4}\\
\hline
\multicolumn{1}{|c|}{\bf Liquid} & \multicolumn{1}{|c|}{7.2} &  \multicolumn{1}{|c|}{13.3} &  \multicolumn{1}{|c|}{9.9} &  \multicolumn{1}{|c|}{8.5} &  \multicolumn{1}{|c|}{\bf 55.9}&  \multicolumn{1}{|c|}{5.2}\\
\hline
\multicolumn{1}{|c|}{\bf Nasal} & \multicolumn{1}{|c|}{0.5} &  \multicolumn{1}{|c|}{1.9} &  \multicolumn{1}{|c|}{3.4} &  \multicolumn{1}{|c|}{1.5} &  \multicolumn{1}{|c|}{7.9}&  \multicolumn{1}{|c|}{\bf 84.8}\\
\hline
\end{tabular}}
}
\end{table}  

To further demonstrate the ability of the proposed features for discriminating different sonorant classes, in addition to MFCC, two SVM classifiers (one using sonority feature and the other using MFCC feature) are fused at score level~\cite{kittler1998combining}. 
For this thirteen-dimensional MFCC feature is used to develop another six class SVM classifier (with RBF kernel, $c=2,\gamma=4$), where $c$ and $\gamma$ values are set using train-test 5-fold cross validation for entire TIMIT test database. For the optimized values of $c$ and $\gamma$, the six-class SVM model is trained. The randomly chosen $80\%$ of TIMIT-test data is used for training and rest $20\%$ is used for testing. The average accuracy of the MFCC based classifier is found to be $80.41\%$. The detailed performance for each class can be seen in Table~\ref{comparision1} (within braces). As there are $6$ classes, each of the classifiers using MFCC and sonority feature will produce $6$ posterior probabilities for each feature vector. 

For the sonority based classifier, the posterior probability scores corresponding to epochs within one frame are averaged to derive single probability score corresponding to each class for each frame. The mean value of probabilities of the two classifiers for each class corresponding to each frame is calculated to derive the fused probability score. The class with maximum average probability score is considered as final output of the combined classifier. The resultant accuracy of the combined classifier is found to be $84.51\%$, which is $80.41\%$ when only MFCC feature is used. The classification accuracy for each class using the combined classifier and only MFCC based classifier is shown in Table~\ref{comparision1} for comparison. By comparing both \% accuracy values in Table~\ref{comparision1}, an absolute improvement of $4.1\%$ can be observed when the two classifiers are fused. For each of the classes, along with improvement in classification, reduction in confusion among different sonorant classes can 
also be observed. It is interesting to observe from Table~\ref{comparision1} that, with increase in correct classification of each class, the percentage of confusion with other classes is reduced for most of the cases.
 
 \begin{table*} [th]
\caption{\label{comparision1} {\textit{Classification accuracy of different sonorant segments (frame level) from TIMIT database using combined sonority and MFCC feature based SVM classifier. Classification accuracy obtained by using only MFCC feature vector is shown within braces ($c=2,\gamma=4$)}}}
\renewcommand{\arraystretch}{1.2}
\vspace{-.2cm}
\centerline{
\scalebox{0.95}{
%\centerline{
\begin{tabular}{|c c c c c c c|}
\hline
\multicolumn{1}{|c|}{\bf Category} & \multicolumn{6}{|c|}{\bf \% Accuracy}\\
\cline{2-7}
\multicolumn{1}{|c|}{\bf } & \multicolumn{1}{|c|}{\bf Low-vowel} &  \multicolumn{1}{|c|}{ \bf Mid-vowel} &  \multicolumn{1}{|c|}{ \bf High-vowel} &  \multicolumn{1}{|c|}{ \bf Glide} &  \multicolumn{1}{|c|}{ \bf Liquid}&  \multicolumn{1}{|c|}{ \bf Nasal}\\
\hline
\hline
\multicolumn{1}{|c|}{\bf Low-vowel} & \multicolumn{1}{|c|}{\bf 86.3 (78.7)} &  \multicolumn{1}{|c|}{6.5 (13.6)} &  \multicolumn{1}{|c|}{3.2 (3.4)} &  \multicolumn{1}{|c|}{0.2 (0.4)} &  \multicolumn{1}{|c|}{3.8 (2.7)}&  \multicolumn{1}{|c|}{0.0 (0.9)}\\
\hline
\multicolumn{1}{|c|}{\bf Mid-vowel} & \multicolumn{1}{|c|}{10.8 (10.4)} &  \multicolumn{1}{|c|}{ \bf 75.4 (68.8)} &  \multicolumn{1}{|c|}{5.3 (11.4)} &  \multicolumn{1}{|c|}{0.7 (1.6)} &  \multicolumn{1}{|c|}{7.8 (7.3)}&  \multicolumn{1}{|c|}{0.0 (0.5)}\\
\hline
\multicolumn{1}{|c|}{\bf High-vowel} & \multicolumn{1}{|c|}{0.5 (0.4)} &  \multicolumn{1}{|c|}{7.3 (9.6)} &  \multicolumn{1}{|c|}{\bf 85.2 (80.8)} &  \multicolumn{1}{|c|}{5.8 (5.1)} &  \multicolumn{1}{|c|}{0.9 (3.1)}&  \multicolumn{1}{|c|}{0.3 (0.4)}\\
\hline
\multicolumn{1}{|c|}{\bf Glide} & \multicolumn{1}{|c|}{0.1 (0.3)} &  \multicolumn{1}{|c|}{2.0 (1.1)} &  \multicolumn{1}{|c|}{6.8 (8.8)} &  \multicolumn{1}{|c|}{ \bf 83.5 (80.7)} &  \multicolumn{1}{|c|}{5.4 (5.3)}&  \multicolumn{1}{|c|}{2.2 (3.8)}\\
\hline
\multicolumn{1}{|c|}{\bf Liquid} & \multicolumn{1}{|c|}{3.6 (4.1)} &  \multicolumn{1}{|c|}{6.5 (5.3)} &  \multicolumn{1}{|c|}{1.5 (2.8)} &  \multicolumn{1}{|c|}{5.8 (4.5)} &  \multicolumn{1}{|c|}{\bf 80.7 (78.8)}&  \multicolumn{1}{|c|}{1.9 (4.5)}\\
\hline
\multicolumn{1}{|c|}{\bf Nasal} & \multicolumn{1}{|c|}{0.2 (0.2)} &  \multicolumn{1}{|c|}{0.8 (0.5)} &  \multicolumn{1}{|c|}{0.8 (1.8)} &  \multicolumn{1}{|c|}{0.9 (1.2)} &  \multicolumn{1}{|c|}{1.3 (1.7)}&  \multicolumn{1}{|c|}{\bf 96.0 (94.7)}\\
\hline
\end{tabular}}
}
\end{table*}

To study individual performances of sonorant classification for male and female, we have developed two sonorant classifiers using SVM (with RBF kernel, $c = 256$, $\gamma = 16$) for male and female utterances from TIMIT test database. For developing each classifier $80\%$ of male/female data is used for training and rest $20\%$ is used for testing. The average accuracy of the six class sonorant classification is found to be is $68.4\%$ for male and $65.6\%$  for female. The relatively poor performance for the female case may be attributed to the associated high non-stationarity nature. 

\subsection{Effect of noise on sonority feature}
\begin{figure}[h]
\centering
\centerline{\epsfig{figure=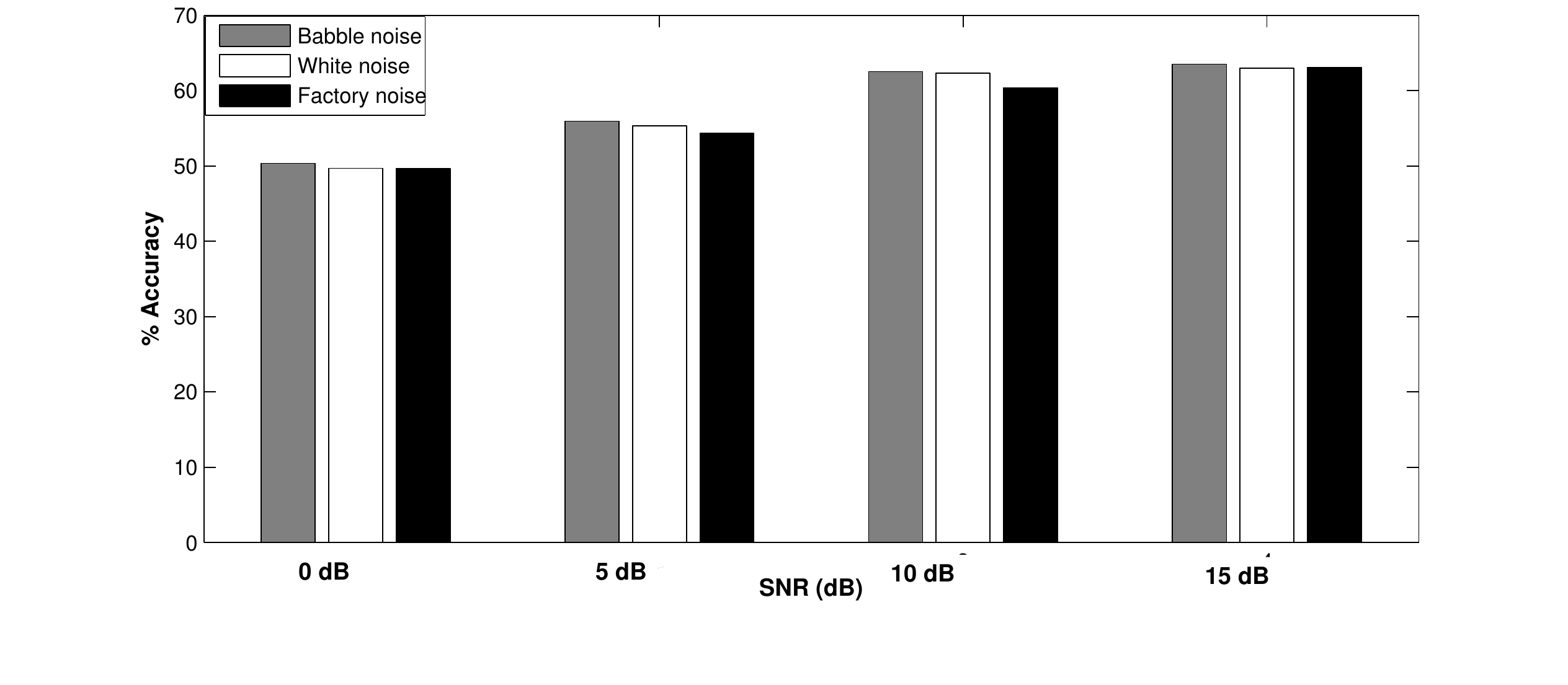,height=1.8in, width=3.5in}}
\vspace{-.5cm}
\caption{{\it Bar plot representing average \% accuracy for SVM based six-class sonorant segment classification in presence of different types of noise with different SNR levels.}}
\label{noise}
\end{figure}

In order to analyze the impact of noise on the proposed features, the classifier trained using features derived from clean speech is employed for testing of noisy cases. The test features are derived after addition of different kinds of noises (babble noise, factory noise, white noise) to the speech signal at different SNR levels ($0$ dB, $5$ dB, $10$ dB, $15$ dB). The average accuracy the six classes for different types and levels of noise is shown as bar plot in Fig.~\ref{noise}. It can be observed that \% accuracy significantly decreases in case of $0$ dB and $5$ dB SNR levels. Whereas, for $10$ dB and $15$ dB cases, \% accuracy is less effected. Further, to analyze the robustness of each of the system, source and suprasegmental features, three six-class SVM classifiers are developed using individual features derived from clean speech. The test features are derived after adding different levels of babble noise with the speech signal. 

\begin{figure}[h]
\centering
\centerline{\epsfig{figure=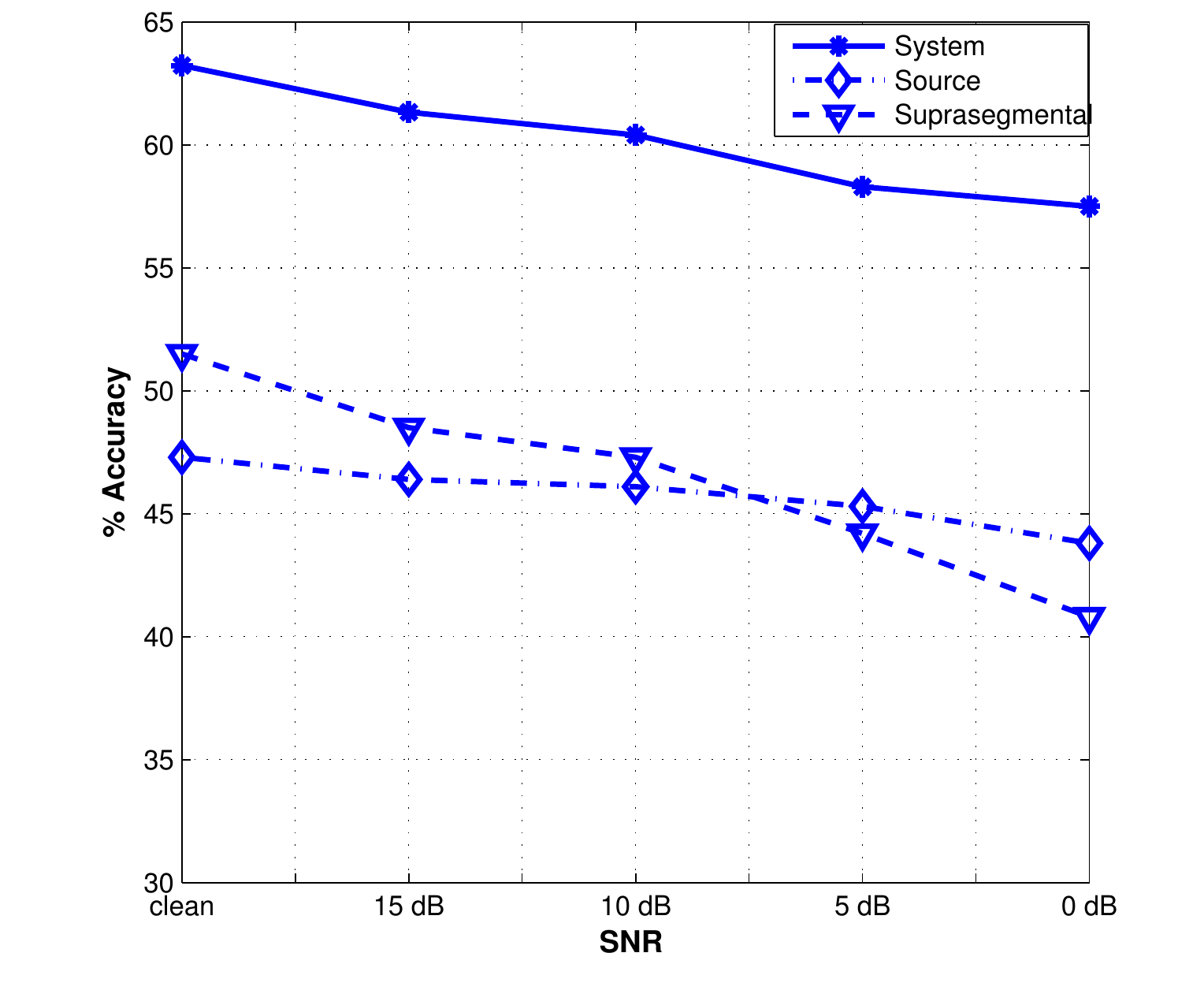,height=2.2in, width=3.1in}}
\vspace{-.5cm}
\caption{{\it Average \% accuracy of six-class sonorant classifier using each of the system, source and suprasegmental features in with respect to different levels of noise.}}
\label{noise2}
\end{figure}

Figure~\ref{noise2} demonstrates degradation of $\%$ accuracy of the three classifiers with increased noise level. This depicts that the suprasegmental feature is more affected due to noise compared to the features of vocal-tract system and excitation source. This may be due to the reason that, suprasegmental feature is directly derived from the speech signal by measuring correlation over successive pitch periods. Furthermore, it is not derived in synchrony with glottal closed phase which may be less susceptible to degradation due to noise. The features of vocal-tract system are derived from HNGD spectrum which is reported to be less affected by different types of noise~\cite{bayya2013spectro}. This happens due to the short and tapered window used in HNGD. For deriving feature of excitation source, the samples corresponding to glottal closed phase around epoch locations is accessed. Hence this feature is also found to be less affected by noise.
  
%%%%%%%%%%%%%%%%%%%%%%%%%%%%%%%%%%%%%%%%%%%%%%%%%%%%%%%%%%%%%%%%%%%%%%%%%%%%%%%%%%%%%%%%%%%%%%%%%%%%%%%%%%%%%%%%%%%%%%%%%%%%%%%%%%%%%%%

The above experiments validate the effectiveness of the proposed feature in discriminating the sonorant sounds or characterization of degree of sonority from given speech signal. To show its usefulness in different speech processing applications, the proposed sonority feature is used in phoneme recognition.

\subsection{Sonority as a feature for phoneme recognizer}
The proposed sonority feature may also be helpful to improve the performance of a phoneme recognizer by incorporating additional information to reduce confusion among different sonorants. In this regard, phoneme recognition framework for TIMIT database is developed in Kaldi toolkit~\cite{kaldi_address,Povey_ASRU2011_kaldi}, where deep neural network (DNN) based acoustic modeling is implemented~\cite{DNN_Hinton_SP_mag}. In addition to traditional MFCC feature, proposed seven-dimensional weighted sonority feature is employed for developing the recognizer. The proposed feature is epoch synchronous. In order to match dimension with MFCC at frame level, average value of feature corresponding to epochs within one frame is calculated. It is then appended with the thirteen-dimensional MFCC feature resulting in a twenty-dimension feature vector. A bigram phoneme language model created from the training set is incorporated in the recognizer.

The $61$ phonemes are mapped into $39$ phonemes for training and testing, the acoustic model is an HMM-DNN hybrid model. The training set contains $3,696$ sentences from $462$ speakers. The development set contains $400$ sentences from $50$ speakers. Core test set is also used as test set, which contains $192$ sentences from $24$ speakers. The number of hidden layers used is $2$. It is reported in Kaldi documentation that $4$ hidden layers are effective when $100$ hours of speech data is available. An initial learning rate of $0.015$ is selected which is reduced to $0.002$ in $20$ epochs. Additional $10$ epochs are employed after reducing the learning rate to $0.002$. Kaldi employs a preconditioned form of stochastic gradient descent. A matrix-valued learning rate is employed instead of using a scalar learning rate in order to reduce the learning rate in dimensions where the derivatives have a high variance. This is in order to control 
instability and stop the parameters moving too fast in any one direction. 

\begin{figure}[h]
\centering
\centerline{\epsfig{figure=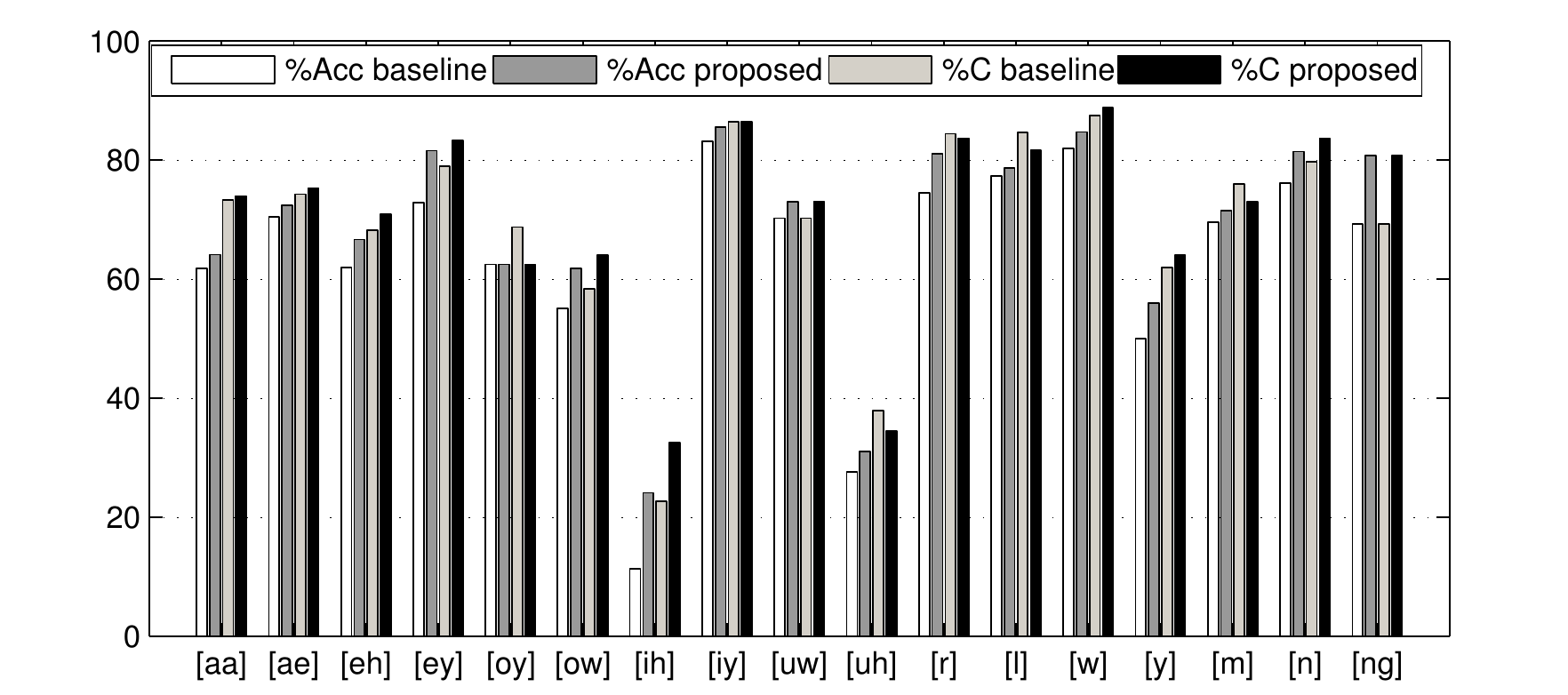,height=1.5in, width=3.8in}}
\vspace{-.3cm}
\caption{{\it Correction percentage (\%C) and accuracy (\%Acc), before and after appending the sonority for various sonorant phones of TIMIT.}}
\label{bar_phonetic1}
\end{figure}

The overall performance of the baseline phoneme recognizer using MFCC as feature and using additional proposed feature (MFCC + sonority) is shown in Table~\ref{phonetic1} in terms phone error rate (\% PER). It is improved while using proposed features along with MFCC. Also, the improvement in case of different sonorant phones in terms of accuracy (\%) and correct (\%) identification is shown in the bar plot of Fig.~\ref{bar_phonetic1}. The performance increases after using the proposed sonority features. It is observed that with the addition of proposed evidence, insertion and substitution of sonorant phones decreases significantly, whereas the reduction in deletion is comparatively less. However, the confusion among different classes of sonorant phones is analyzed in terms of \% substitution. It seems to reduce while employing the proposed feature in addition to MFCC as shown in Table~\ref{phoneme_comparision1}. 

\begin{table} [h]
\caption{\label{phonetic1} {\textit{Phone error rate (PER) for DNN based phoneme recognizer by using MFCC and (MFCC+Sonority) feature}}}
\renewcommand{\arraystretch}{1.4}
\vspace{-2mm}
\centerline{
\scalebox{1}{
%\centerline{
\begin{tabular}{|c c c |}
\hline
\multicolumn{1}{|c|}{\bf{Evaluation on}} & \multicolumn{2}{|c|}{\bf PER(\%)}\\
\cline{2-3}
\multicolumn{1}{|c|}{\bf{}} & \multicolumn{1}{|c|}{\bf MFCC} & \multicolumn{1}{|c|}{\bf MFCC+sonority feature}\\
\hline
\hline
\multicolumn{1}{|c|}{\bf Test set} & \multicolumn{1}{|c|}{22.7} & \multicolumn{1}{|c|}{21.4}\\
\hline
\multicolumn{1}{|c|}{\bf Dev set} & \multicolumn{1}{|c|}{21.2} & \multicolumn{1}{|c|}{20.3}\\
\hline
\end{tabular}}
}
\end{table}

 \begin{table} [th]
\caption{\label{phoneme_comparision1} {\textit{\% substitution of different sonorant phones before and after appending the proposed sonority evidence for various sonorant phones of TIMIT. Baseline result using MFCC is shown braces. }}}
\renewcommand{\arraystretch}{2}
\vspace{-2mm}
\centerline{
\scalebox{0.63}{
%\centerline{
\begin{tabular}{|c c c c c c c c|}
\hline
\multicolumn{1}{|c|}{\bf Category} & \multicolumn{7}{|c|}{\bf \% Substitution}\\
\cline{2-8}
\multicolumn{1}{|c|}{\bf } & \multicolumn{1}{|c|}{\bf Low-vowel} &  \multicolumn{1}{|c|}{ \bf Mid-vowel} &  \multicolumn{1}{|c|}{ \bf High-vowel} &  \multicolumn{1}{|c|}{ \bf Glide} &  \multicolumn{1}{|c|}{ \bf Liquid}&  \multicolumn{1}{|c|}{ \bf Nasal} &  \multicolumn{1}{|c|}{ \bf Total}\\
\hline      
\hline
\multicolumn{1}{|c|}{\bf Low-vowel} & \multicolumn{1}{|c|}{ 4.0 (4.1)} &  \multicolumn{1}{|c|}{5.5 (6.1)} &  \multicolumn{1}{|c|}{6.8 (7.0)} &  \multicolumn{1}{|c|}{0.0 (0.1)} &  \multicolumn{1}{|c|}{0.5 (0.5)}&  \multicolumn{1}{|c|}{0.3 (0.4)} &  \multicolumn{1}{|c|}{\bf 17.1 (18.2)}\\
\hline
\multicolumn{1}{|c|}{\bf Mid-vowel} & \multicolumn{1}{|c|}{3.5 (4.9)} &  \multicolumn{1}{|c|}{ 1.3 (1.3)} &  \multicolumn{1}{|c|}{4.8 (4.9)} &  \multicolumn{1}{|c|}{0.0 (0.0)} &  \multicolumn{1}{|c|}{0.8 (1.2)}&  \multicolumn{1}{|c|}{0.1 (0.1)} &  \multicolumn{1}{|c|}{\bf 10.5 (12.4)}\\
\hline
\multicolumn{1}{|c|}{\bf High-vowel} & \multicolumn{1}{|c|}{4.3 (4.7)} &  \multicolumn{1}{|c|}{1.9 (2.1)} &  \multicolumn{1}{|c|}{ 4.7 (5.3)} &  \multicolumn{1}{|c|}{0.4 (0.9)} &  \multicolumn{1}{|c|}{0.2 (0.2)}&  \multicolumn{1}{|c|}{0.2 (0.7)} &  \multicolumn{1}{|c|}{\bf 11.7 (13.9)}\\
\hline
\multicolumn{1}{|c|}{\bf Glide} & \multicolumn{1}{|c|}{0.0 (0.0)} &  \multicolumn{1}{|c|}{0.0 (0.0)} &  \multicolumn{1}{|c|}{1.0 (1.8)} &  \multicolumn{1}{|c|}{ 0.3 (0.3)} &  \multicolumn{1}{|c|}{0.3 (0.5)}&  \multicolumn{1}{|c|}{0.5 (0.8)} &  \multicolumn{1}{|c|}{\bf 2.1 (3.4)}\\
\hline
\multicolumn{1}{|c|}{\bf Liquid} & \multicolumn{1}{|c|}{0.6 (1.2)} &  \multicolumn{1}{|c|}{0.8 (0.9)} &  \multicolumn{1}{|c|}{0.3 (0.2)} &  \multicolumn{1}{|c|}{0.1 (0.3)} &  \multicolumn{1}{|c|}{ 0.1 (0.1)}&  \multicolumn{1}{|c|}{0.4 (0.7)} &  \multicolumn{1}{|c|}{\bf 2.3 (3.4)}\\
\hline
\multicolumn{1}{|c|}{\bf Nasal} & \multicolumn{1}{|c|}{0.5 (0.5)} &  \multicolumn{1}{|c|}{0.2 (0.3)} &  \multicolumn{1}{|c|}{0.5 (0.5)} &  \multicolumn{1}{|c|}{0.1 (0.2)} &  \multicolumn{1}{|c|}{0.2 (0.3)}&  \multicolumn{1}{|c|}{ 3.6 (3.9)} &  \multicolumn{1}{|c|}{\bf 5.1 (5.7)}\\
\hline
\end{tabular}}
}
\end{table}

\section{Summary, Conclusions and Scope} \label{conclusion}
In this work, an effort is made to define a feature which can represent the degree of sonority associated with a sound unit. For this task, different characteristics of sonorant sounds reflected in the speech signal are analyzed. Consequently features based on vocal-tract system, excitation source and suprasegmental aspects are derived. These features correlate with less vocal-tract constriction, glottal vibration and periodicity properties of sonorant sounds. To justify, whether each of the proposed features can represent the level of sonority, distributions for feature values are shown for different sonorant sounds along the sonority hierarchy. Each of the proposed features shows increasing/decreasing trend in feature value with the increase in sonority. The proposed seven-dimensional sonority feature is used in classification among different sonorant sounds and is found to be potential for the same. It is also shown to be useful for the phoneme recognition application. In future we may focus on exploring 
evidences which can reduce the confusion among adjacent classes in the sonority hierarchy.

\section{Acknowledgement}
\vspace{-1mm}
  This work is a part of the ongoing project on the ``Development of Text-to-Speech Synthesis for Assamese and Manipuri languages'' funded by TDIL, DEiTy, MCIT, GOI. The authors would also like to thank Mr. Abhishek Dey for his kind help in developing DNN based phoneme recognition framework.

{
\bibliographystyle{IEEEtran}
% \baselineskip 14pt   %12, 18 and 24
% \small
% \bibliographystyle{IEEEtran}
% \bibliography{spkr_vlr_ref}
%\bibliographystyle{IEEEtran}
\bibliography{sonority_ref}
}

\begin{IEEEbiography}[{\includegraphics[width=1in,height=1.25in,clip,keepaspectratio]{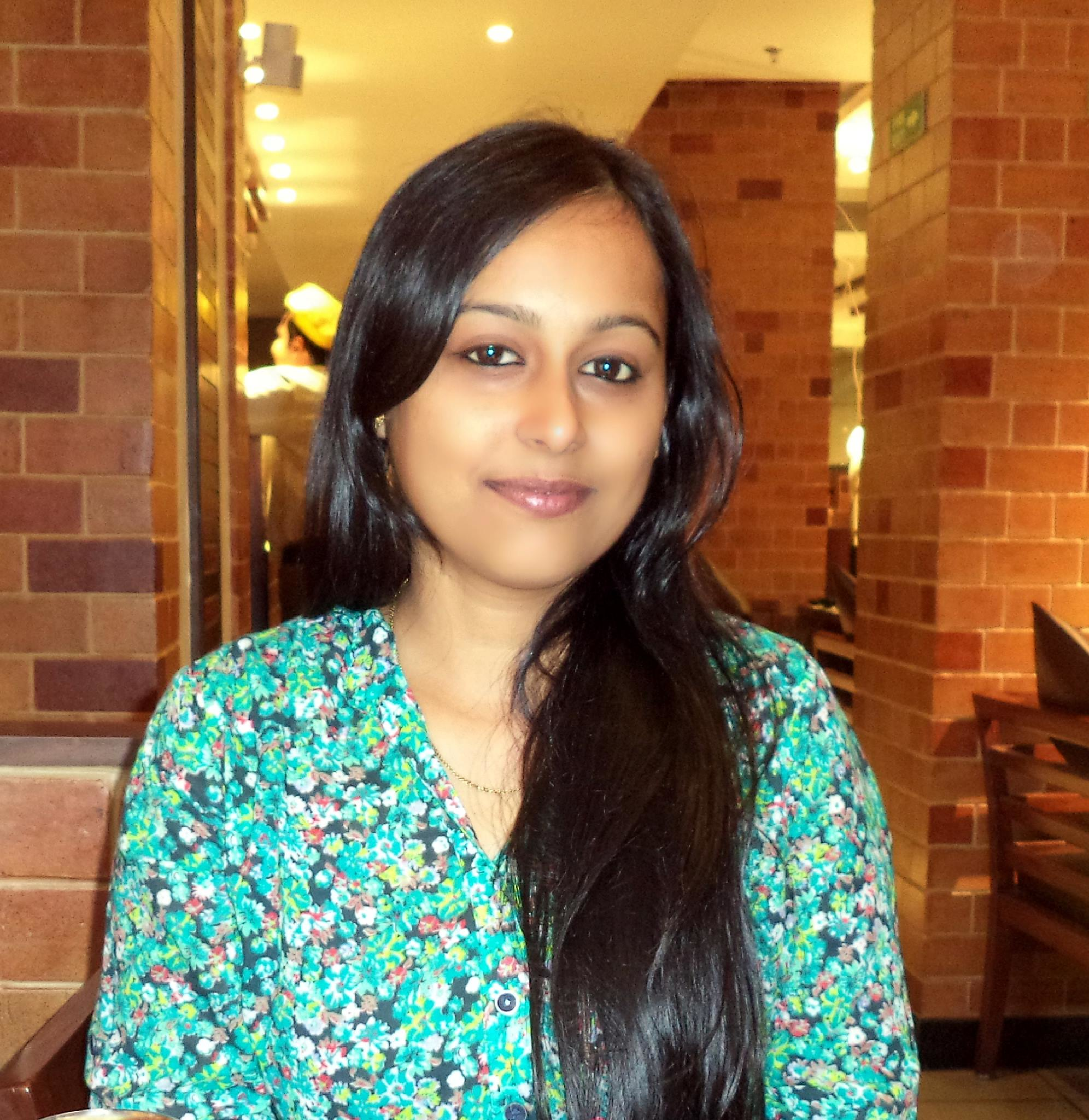}}]{Bidisha Sharma}
received B.E.
degree in Electronics and Telecommunication Engineering from Girijananda Chowdhury 
Institute of Management and Technology (GIMT), Gauhati University,
Guwahati, India, in 2012. She is currently pursuing PhD
in the Department of Electronics and Electrical Engineering,
Indian Institute of Technology (IIT) Guwahati.
Her research interests are in speech signal processing, speech synthesis and speech enhancement.
\end{IEEEbiography}

 \begin{IEEEbiography}[{\includegraphics[width=1in,height=1.25in,clip,keepaspectratio]{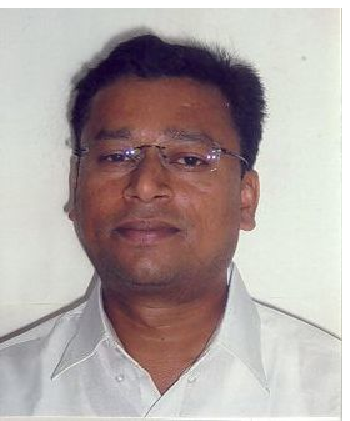}}]{S. R. Mahadeva Prasanna}
received the B.E.
degree in Electronics Engineering from Sri Siddartha
Institute of Technology, Bangalore University,
Bangalore, India, in 1994, the M.Tech. degree in
Industrial Electronics from the National Institute
of Technology Karnataka (NITK), Surathkal, India, in 1997, and the
Ph.D. degree in Computer Science and Engineering
from the Indian Institute of Technology (IIT) Madras,
Chennai, India, in 2004. He is currently a Professor
in the Department of Electronics and Electrical Engineering,
Indian Institute of Technology (IIT) Guwahati.
His research interests are in speech, handwriting and signal processing.
\end{IEEEbiography}
\vspace{-1cm}
%\begin{IEEEbiography}[{\includegraphics[width=1in,height=1.25in,clip,keepaspectratio]{./photo/rohit_sinha}}]{Rohit Sinha}

%\end{IEEEbiography}
\vfill
\vfill

% that's all folks
\end{document}